\documentclass[11pt]{article}

\usepackage{amsmath}
\usepackage{mathrsfs}
\usepackage{graphicx}
\usepackage{amssymb}
\usepackage{dsfont}
\usepackage{a4}
\usepackage{url}
\usepackage{algorithm}
\usepackage{algorithmicx}
\usepackage{algpseudocode}
\usepackage{color}
\usepackage[bookmarks, colorlinks=true, plainpages = false, linkcolor = blue, citecolor = green, urlcolor = blue, filecolor = blue]{hyperref}

\newcommand{\bea}{\begin{eqnarray*}}
\newcommand{\eea}{\end{eqnarray*}}
\newcommand{\be}{\begin{eqnarray}}
\newcommand{\ee}{\end{eqnarray}}

\def\Arg{\mathrm{Arg}}
\def\dd{\mathrm{d}}

\def\diam{\mathrm{diam}}
\def\Pr{\mathrm{Prob}}

\def\vol{\mathrm{vol}}

\DeclareMathOperator{\CR}{\mathsf{CR}}
\DeclareMathOperator{\PR}{\mathsf{PR}}

\def\IBq{I_{B,q}}

\def\ma{\alpha}

\def\mg{\gamma}

\def\Ex{\mathsf{E}}

\def\ra{\rightarrow}

\def\e1{\mathsf{e}}

\def\Ab{\mathbf{A}}
\def\Bb{\mathbf{B}}
\def\Db{\mathbf{D}}

\def\Hb{\mathbf{H}}

\def\Sb{\mathbf{S}}

\def\Xb{\mathbf{X}}
\def\Zb{\mathbf{Z}}

\def\0b{\mathbf{0}}
\def\1b{\mathbf{1}}

\def\cb{\mathbf{c}}

\def\xb{\mathbf{x}}
\def\yb{\mathbf{y}}
\def\zb{\mathbf{z}}

\def\SA{{\mathscr A}}
\def\SB{{\mathscr B}}
\def\SC{{\mathscr C}}
\def\SH{{\mathscr H}}

\def\SL{{\mathscr L}}

\def\SO{\mathcal{O}}

\def\SX{{\mathscr X}}

\newcommand{\fin}{\mbox{}~\hfill\mbox{$\Box$}}
\newcommand{\carre}{\mbox{}~\hfill\rule{2mm}{2mm}}

\newtheorem{remark}{Remark}
\newtheorem{definition}{Definition}
\newtheorem{theorem}{Theorem}

\newcommand{\vsp}{\vspace{0.3cm}}

\begin{document}

\title{Incremental space-filling design based on coverings and spacings: improving upon low discrepancy sequences}

\author{A. Nogales G\'omez, L. Pronzato, M.-J. Rendas \\
\mbox{}\\
        UCA, CNRS, Laboratoire I3S, UMR 7271,  \\
        B\^at.\ Euclide, Les Algorithmes, 2000 route des lucioles,\\
        06900 Sophia Antipolis cedex, France \\
              {\tt \{amaya.nogales-gomez,pronzato,rendas\}@i3s.unice.fr}           
}

\date{\today}

\maketitle

\begin{abstract} The paper addresses the problem of defining families of ordered sequences $\{\xb_i\}_{i\in \mathds{N}}$ of elements of a compact subset $\SX$ of $\mathds{R}^d$ whose prefixes $\Xb_n=\{\xb_i\}_{i=1}^{n}$, for all orders $n$, have good space-filling properties as measured by the dispersion (covering radius) criterion. Our ultimate aim is the definition of incremental algorithms that generate sequences $\Xb_n$ with small optimality gap, i.e., with a small increase in the maximum distance between points of $\SX$ and the elements of $\Xb_n$ with respect to the optimal solution $\Xb_n^\star$. The paper is a first step in this direction, presenting incremental design algorithms with proven optimality bound for one-parameter families of criteria based on coverings and spacings that both converge to dispersion for large values of their parameter. The examples presented show that the covering-based method outperforms state-of-the-art competitors, including coffee-house, suggesting that it inherits from its guaranteed 50\% optimality gap.

{\bf keywords} {Covering; Spacing; Submodularity; Greedy algorithm; Computer experiments; Space-filling design}

\end{abstract}

\section{Introduction and motivation}

The paper discusses algorithmic constructions to define space-filling designs. Given
a compact subset $\SX\subset \mathds{R}^d$, we say that a finite subset $\Zb_n\subset \SX$ is a space-filling design if $\Zb_n$ fills $\SX$ evenly. Several mathematical definitions of this intuitive notion have been proposed in the literature, and we refer the interested reader to the reviews \cite{P-JSFdS2017,PM_SC_2012} and the books \cite{FangLS2006}, \cite[Chap.~5]{SantnerWN2003} for a comprehensive presentation and discussion. Generically, a space-filling criterion is a set function, that maps (possibly finite) subsets of $\SX$ to $\mathbb{R}$.

In this work, the space-filling quality of a design $\Zb_n$ is measured through the distances from the domain points to their closest point in $\Zb_n$, i.e., their distance to the design. Motivated by interpolation problems, the covering radius, equal to the maximum of these distances, is often considered to be the ultimate measure of the space-filling quality of $\Zb_n$: good space-filling sequences should have small covering radius. In colloquial terms, ``good designs leave no large holes in $\SX$''.

The packing radius is in some sense a dual space-filling metric, assessing how well $\Zb_n$ is spread inside $\SX$ through the distance between its closest points. In terms of this criterion, a design is space-filling if it has a large packing radius. It is easy to see that the designs optimal for the covering radius are not optimal according to the packing radius: the latter must have some points in the boundary of $\SX$ (otherwise a larger packing radius would be obtained by expanding the design uniformly), while this clearly does not lead to minimal covering radius. Nevertheless, the packing radius is widely used as a space-filling criterion, given the much smaller  numerical complexity involved in its computation when compared to the covering radius, and in the paper we also consider the optimization of the packing radius.

Both the covering and packing radii have important limitations as space-filling criteria, since they only reflect local characteristics. Indeed, they are generally determined by a few points of $\Zb_n$, being insensitive to perturbations of the other points, with the consequence that many extensions of a given design may keep these indicators constant. Moreover, when $d$ is large the numerical estimation of the covering radius is difficult and, as we shall see, very sensitive to the presence of vertices in $\SX$. As already discussed in \cite{NoonanZ2020}, the $\ma$-quantile of the distribution of the distance of a random point of $\SX$ to $\Zb_n$, which we shall call the covering $\ma$-quantile of the design, with $\ma$ close to 1, is a more stable space-filling indicator. For completeness, we will also compare designs according to this performance measure.

Most literature on space-filling design considers the construction of designs of a given target size. We are instead interested in the more complex situation where the number of design points that will effectively be used, $n\leq n_{\max}$, is to be decided in a later step, and our goal is to specify an ordered design $\Xb_{n_{\max}}$ such that all prefix designs $\Xb_n=\{\xb_i\}_{i=1}^n$, $n\leq n_{\max}$, have good space-filling properties. Such a situation typically occurs in the construction of the initial design on which the metamodel subsequently used in a computer experiment will be identified. In this context, performance is no longer assessed through a single scalar, but rather by the trajectory of criterion values for increasing design sizes.

\paragraph{Background.}
Optimization of the covering radius is a  highly non-linear (the cost function involves a maximum) and inherently multi-dimensional (the search space has dimension $n\times d$) combinatorial problem. Actually, the problem is known to be NP-hard, see \cite[p.~414]{KorteV2012}, meaning that except for toy-problems we can only hope to find tractable algorithms that produce reasonably good solutions. In these circumstances, it is important to know how far the solutions found by a given algorithm can be from optimality, commonly designated in algorithmics by optimality gap. The definition of algorithms with guaranteed optimality gap for NP-hard problems is an active research topic which has produced important achievements for many combinatorial optimization problems, and our work is a contribution in this sense.

Algorithms which  define the solution by incrementally adding a point at a time are specially interesting given their more manageable complexity: at each iteration an optimization problem in $\SX$, i.e., in only $d$ variables, needs to be solved. It is a well known --- and easily verified --- fact that besides additive cost functions, for which the optimal solution can be found by greedily appending the point in $\SX$ that produces the largest improvement, the optimality gap of greedy algorithms can be bounded if the criterion optimized is a submodular set function (this notion will be made precise below). Unfortunately, neither the covering radius nor the packing radius are submodular set functions.

\paragraph{Claims and hint of the contents.}

In this paper, we present a parameterized submodular design criterion, based of the cumulative distribution function (c.d.f.) of the distance of a random point to the design, which is asymptotically (for large values of its parameter) related to the covering radius, submodularity implying that its greedy solutions have a bounded optimality gap. We show numerically that the performance of the designs produced is robust with respect to the precise tuning parameters of the criterion. Interestingly, our construction is free from knowledge of the detailed geometry of the domain $\SX$, requiring only the specification of two (possibly coincident) sets: a set of eligible design points and a set of points used to compute the c.d.f., which must finely cover $\SX$.

Despite its simplicity, the incremental construction of designs by greedy maximization of the packing radius, sometimes called coffee-house design in the literature \cite{Muller2000,Muller2007}, ideally ensures an optimality gap of 50\% for the covering radius \cite{Gonzalez85}. Practical constructions use a finite set of candidates $\SX_C$ at each iteration, and the 50\% guarantee above holds for $\SX_C$ only. As explained in the paper, the quality of the resulting design heavily depends on the imbedding of $\SX_C$ in $\SX$. We show that a slight modification of the original coffee-house algorithm based on the notion of spacings, which keeps the design points away from the boundary of $\SX$ with the aim of reducing the covering radius, improves robustness with respect to the choice of $\SX_C$.

Our numerical studies indicate that the greedy optimization of our c.d.f.-based criterion yields design with better covering properties than all alternative techniques tested: low-discrepancy sequences (LDS), variants of coffee-house design, and incremental constructions based on two other relaxed versions of the covering radius\footnote{There exist other incremental constructions that generate space-filling designs with good covering performance, in particular those based on
maximization of mutual information \cite{BeckG2016,KrauseSG2008} or on minimization of a kernel discrepancy by kernel herding \cite{PZ2020-SIAM}. They rely on the choice of a suitable positive definite kernel and are especially adapted to interpolation based on Gaussian process models. We do not follow this approach here, and the methods we consider are based only on geometrical considerations. A thorough comparison would certainly be of interest, but is beyond the scope of this paper.}. In particular, when the domain is a $d$-dimensional hypercube and performance is measured by the covering radius, it always outperforms the other methods.

\paragraph{Paper organization.}
This paper extends preliminary work presented at the SIAM Conference on Uncertainty Quantification in 2016 \cite{PronzatoR:SIAMUQ2016}.
It is organized as follows. In Section~\ref{S:submodularity} we recall the definition of submodular set functions and the fundamental theorem that establishes a bound on the optimality gap of their greedy solutions. Section~\ref{S:covering} presents the novel c.d.f.-based design criterion, addressing in detail its numerical evaluation and illustrating its behavior for several parametrization choices. Other greedy constructions, based on packing radius and spacings, are presented in Section~\ref{S:CH}. Finally, Section~\ref{S:Examples} presents a comparative performance study of the proposed methods against a set of state-of-the-art space-filling design methods.

\paragraph{Basic definitions and notation.}

Throughout the paper, $\SX$ is a compact subset of $\mathds{R}^d$ with nonempty interior, with the hypercube $\SC_d=[0,1]^d$ as a typical example. $\1b_d$ denotes the $d$-dimensional vector with all components equal to one, and the center of $\SC_d$ is thus $\1b_d/2$.
Let $\Zb_n=\{\zb_1,\ldots,\zb_n\}$ be any $n$-point design in $\SX$. For any point $\xb\in\SX$, we denote $d(\xb,\Zb_n)=\min_{i=1,\ldots,n} \|\xb-\zb_i\|$, with $\|\cdot\|$ the $\ell_2$ norm.
The covering radius $\CR(\Zb_n)$ of $\Zb_n$ is defined by
\bea
\CR(\Zb_n) = \CR_{[\SX]}(\Zb_n) = \max_{\xb\in\SX} d(\xb,\Zb_n) \,.
\eea
It is called dispersion in the theory of quasi-Monte Carlo methods \cite[Chap.~6]{Niederreiter92} and coincides with the minimax-distance, or fill criterion, used in computer experiments and function interpolation; see \cite{DeMarchi2003,JohnsonMY90,P-JSFdS2017,PM_SC_2012}.
When the design objective is to ensure an accurate prediction of the values of an unknown function $f$ over all $\SX$ based on evaluations at $\Zb_n\subset\SX$, it is important to ensure that for any $\xb$ in $\SX$ there always exists a $\zb_i$ at proximity of $\xb$ where $f(\zb_i)$ has been evaluated (see, e.g., \cite{Schaback95} for a precise formulation and error bounds), making designs with a small value of $\CR(\Zb_n)$ particularly desirable.

Let $\SX_N=\{\xb^{(1)}, \xb^{(2)},\ldots,\xb^{(N)}\}$ be a finite subset of $\SX$, with $N\gg n$, that is well spread over $\SX$; it may be a regular grid when $\SX=\SC_d$ and $d$ is small, or (possibly after a suitable rescaling of $\SX$) the first $N$ points of a LDS in $\SC_d$ that fall in $\SX$. With a slight abuse of notation, we assume that $\SX_N$ and $\SX_{N'}$ do not necessarily share any elements when $N \neq N'$. Note that the value
\be\label{CR-approx}
\CR_{[\SX_N]}(\Zb_n) = \max_{\xb\in\SX_N} d(\xb,\Zb_n)
\ee
underestimates $\CR(\Zb_n)$ by an amount depending on how well $\SX_N$ is itself spread in $\SX$. In particular, we have
\be\label{ineq-CR-approx}
\CR_{[\SX_N]}(\Zb_n) \leq \CR(\Zb_n) \leq \max_{\xb\in\SX} \min_{\zb_i\in\Zb_n, \xb^{(j)}\in\SX_N} \left( \|\xb-\xb^{(j)}\|+\|\xb^{(j)}-\zb_i\| \right)  \leq \CR_{[\SX_N]}(\Zb_n) + \CR(\SX_N) \,.
\ee
In the appendix we give an upper bound on $\CR(\SX_N)$ when $\SX_N$ is the prefix of a LDS.

The packing radius (also designated by separating radius, or maximin-distance)
\bea
\PR(\Zb_n) = \min_{\zb_i\neq\zb_j\in\Zb_n} \frac12\, \|\zb_i-\zb_j\| \ (n\geq 2)\,,
\eea
is also often used as a space-filling characteristic of a design $\Zb_n$ \cite{DeMarchi2003,JohnsonMY90,P-JSFdS2017,PM_SC_2012}, in particular due to the simplicity of its calculation when compared to $\CR(\Zb_n)$. Good designs should have a large packing radius. Notice, however, that $\PR(\Zb_n)$ is an even more local space-filling characteristic of $\Zb_n$ than $\CR(\Zb_n)$, in the sense that moving a single point $\zb_i$ of $\Zb_n$ to make it coincide with another $\zb_j$ sets $\PR(\Zb_n)$ to zero whereas $\CR(\Zb_n)$ only increases to $\CR(\Zb_n\setminus\{\zb_i\})$.

We reserve the notation $\Xb_n=[\xb_1,\ldots,\xb_n]$ to designs that are constructed incrementally, our objective being that all prefix designs $\Xb_m$, $m=1,\ldots,n$, should have good space-filling properties.
A special notation is used for the Sobol' and Halton sequences, with $\Sb_{n,d}$ and $\Hb_{n,d}$ respectively denoting their first $n$ points in dimension $d$.

When using Gaussian process modeling and kriging, the number of design points required for a given prediction accuracy can be related to the correlation length of the model \cite{HarariBDH2018,P-RESS2019}. However, this requires prior knowledge on the behavior of the function to be approximated, or that preliminary experimentation be performed with an initial design. When this initial design is suitably ordered, function evaluations can be interrupted when the confidence in the identified model is deemed sufficient to proceed to a second design phase, where complementary design points are added to improve the precision of the model predictions. In this situation, the size $n$ of the initial design is small, much smaller than $2^d$ when $d$ is large, and in particular we can have in mind the ``$10\,d$'' rule of \cite{LoeppkySW2009}.

Interest in incremental constructions brings an important constraint to the optimization problem, since designs with minimum covering radius are not nested. The consequences are well known in the large $n$ situation: for $d=1$, remember for instance the low dispersion sequence of Ruzsa whose discrepancy does not tend to zero as $n\ra \infty$ \cite[p.~154]{Niederreiter92}. But the phenomenon is already present for small $n$: for $\SX=\SC_d$ the one-point $\CR$-optimal design is $\Zb_1^\star=\{\1b_d/2\}$ (the center of $\SC_d$), whereas 2-point $\CR$-optimal designs have the form $\Zb_2^\star=\{\zb_1^\star,\zb_2^\star=\1b_d-\zb_1^\star\}$ with $\zb_1^\star$ having all it coordinates equal to $1/2$ except one which equals $1/4$; see the appendix. Any incremental construction is therefore already suboptimal for $\CR(\Xb_n)$ at $n=2$ (and $\CR(\Xb_n)$ can remain equal to $\sqrt{d}/2$ for some iterations when the first design points are chosen near the vertices of $\SC_d$; see \cite{CabralFPR2020}).

\section{Performance guarantee for submodular function maximization}\label{S:submodularity}
This section recalls the notion of submodular set functions and a well known theorem establishing, when they are non-decreasing, bounds on the optimality gap of their greedy maximization.
\subsection{Submodularity and the greedy algorithm}

Let $\SX_C$ be a finite candidate set with $C=|\SX_C|$ elements and denote by $2^{\SX_C}$ its power set. A (scalar real valued) set-function is an application of $2^{\SX_C}$ on $\mathds{R}$. Space-filling criteria are set-functions. We say that a set-function $f$ is non-decreasing when $f(\Ab\cup\{\xb\}) \geq f(\Ab)$ for any $\Ab\subset\SX_C$ and any $\xb\in\SX_C$.

\begin{definition}{\bf (Submodular function)} \label{Def:1}
A set-function $f: 2^{\SX_C} \ra \mathds{R}$ is submodular if and only if it satisfies the following three equivalent conditions, see e.g., \cite{Bach2013}.
\be
(a) \quad \quad && f(\Ab)+f(\Bb)\geq f(\Ab\cup\Bb)+f(\Ab\cap\Bb), \qquad \forall \Ab, \Bb \in  2^{\SX_C}\,; \nonumber\\
(b) \quad \quad &&f(\Ab\cup\{\xb\})-f(\Ab) \geq f(\Bb\cup\{\xb\})-f(\Bb), \qquad \forall \Ab \subset \Bb \in  2^{\SX_C},\ \xb\in \SX_C\setminus\Bb\,; \label{eq:dimret}\\
(c) \quad \quad &&f(\Ab\cup\{\xb\})-f(\Ab) \geq f(\Ab\cup\{\xb,\yb\})-f(\Ab\cup\{\yb\}), \qquad \forall \Ab, \Bb \in  2^{\SX_C},\ \xb,\yb\in \SX_C\setminus\Ab\,. \nonumber
\ee
\end{definition}
Inequality \eqref{eq:dimret} is known as the {\em diminishing returns} property, stating that the increment resulting of the addition of an element $\xb$ to a set is a decreasing function of the set to which it is added.

The problem of maximizing a set-function is in general NP-hard, its exact resolution requiring the evaluation of $f$ over all the $2^C$ elements of the power set, which is infeasible except in trivial cases of little practical interest. In this paper we concentrate on the greedy (one-step-ahead) algorithm below to efficiently find approximate maximizers of size $k$ of a given set-function $f$.

\begin{algorithm}[ht]
\caption{(Greedy Algorithm)}\label{Algo:greedy}
\begin{algorithmic}[1]
\State set $\Xb=\emptyset$
\While{ $|\Xb|<k$}
\State{find $\xb$ in $\SX_C$ such that $f(\Xb\cup\{\xb\})$ is maximal}
\State{$\Xb \leftarrow \Xb\cup\{\xb\}$}
\EndWhile
\State \textbf{return} $\Xb$
\end{algorithmic}
\end{algorithm}

When several solutions exist at Step~3, a single one is selected (e.g. randomly).
In spite of its simplicity, under mild conditions on $f$ Algorithm~1 can be fairly efficient when applied to a non-decreasing submodular function as stated by the following theorem \cite{NemhauserWF1978}.

\begin{theorem}\label{th:NemhauserWF}
Let $f$ be a non-decreasing submodular set function, then, for any given $k$, $1\leq k \leq C$, Algorithm~1 returns a set $\Xb_k$ with bounded optimality gap
\begin{equation}\label{bound-greedy-1}
\frac{f^\star_k-f(\Xb_k)}{f^\star_k-f(\emptyset)} \leq (1-1/k)^k \leq 1/\e1 <  0.3679 \,,
\end{equation}
where $f^\star_k=\max_{\Xb\subset\SX_C: |\Xb|\leq k} f(\Xb)$ and $\e1=\exp(1)$.
\end{theorem}

Note that \eqref{bound-greedy-1} implies the efficiency bound
\be
\frac{f(\Xb_k)- f(\emptyset)}{f^\star_k - f(\emptyset)} \geq 1-\frac{1}{\e1} > 63.2\%\,, \ \ \forall k\geq 1 \,. \label{*th1}
\ee

\subsection{The lazy-greedy algorithm}\label{S:lazy-greedy}

\begin{figure}[ht!]
\begin{center}
 \includegraphics[width=.49\linewidth]{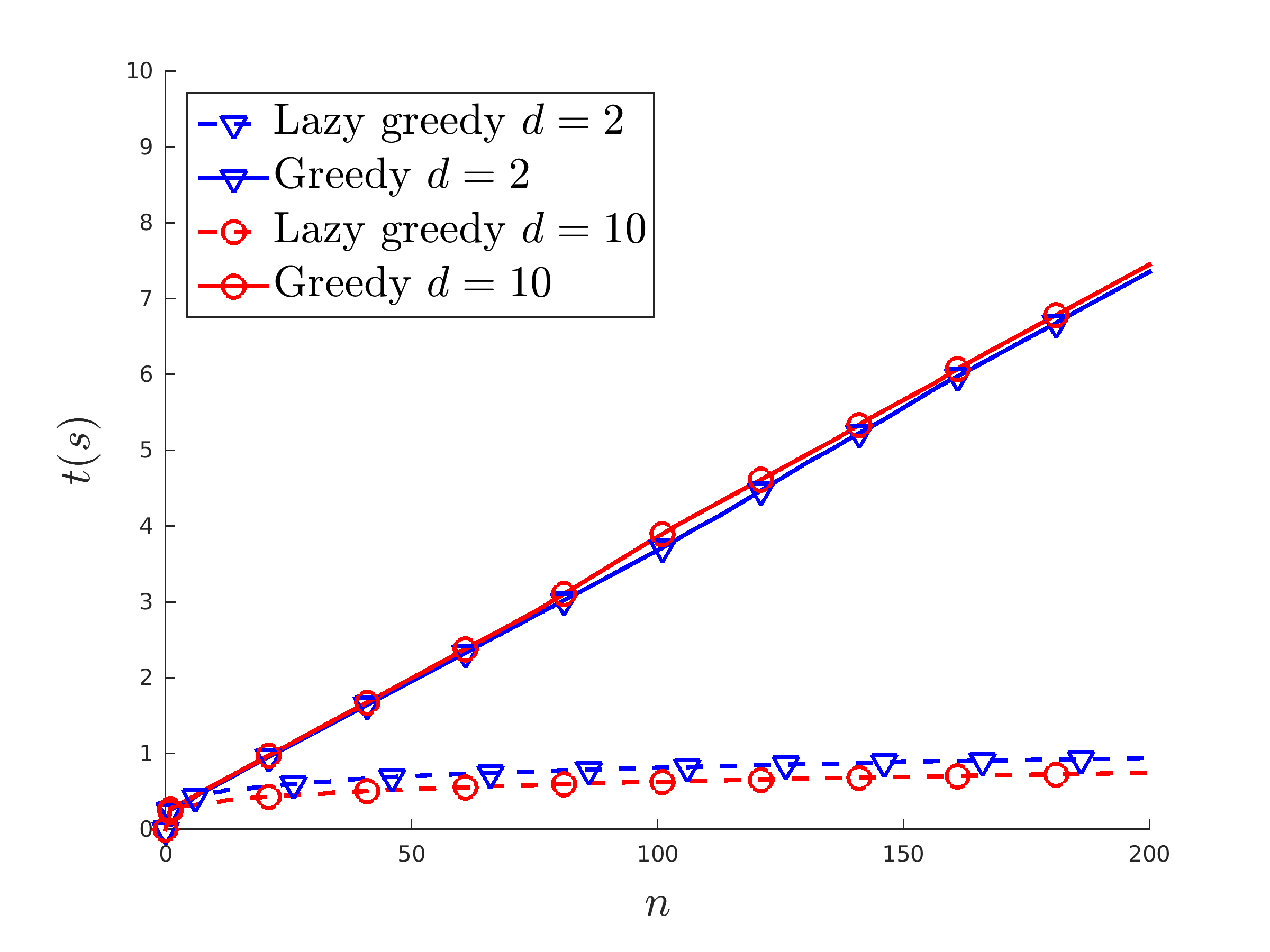} \includegraphics[width=.49\linewidth]{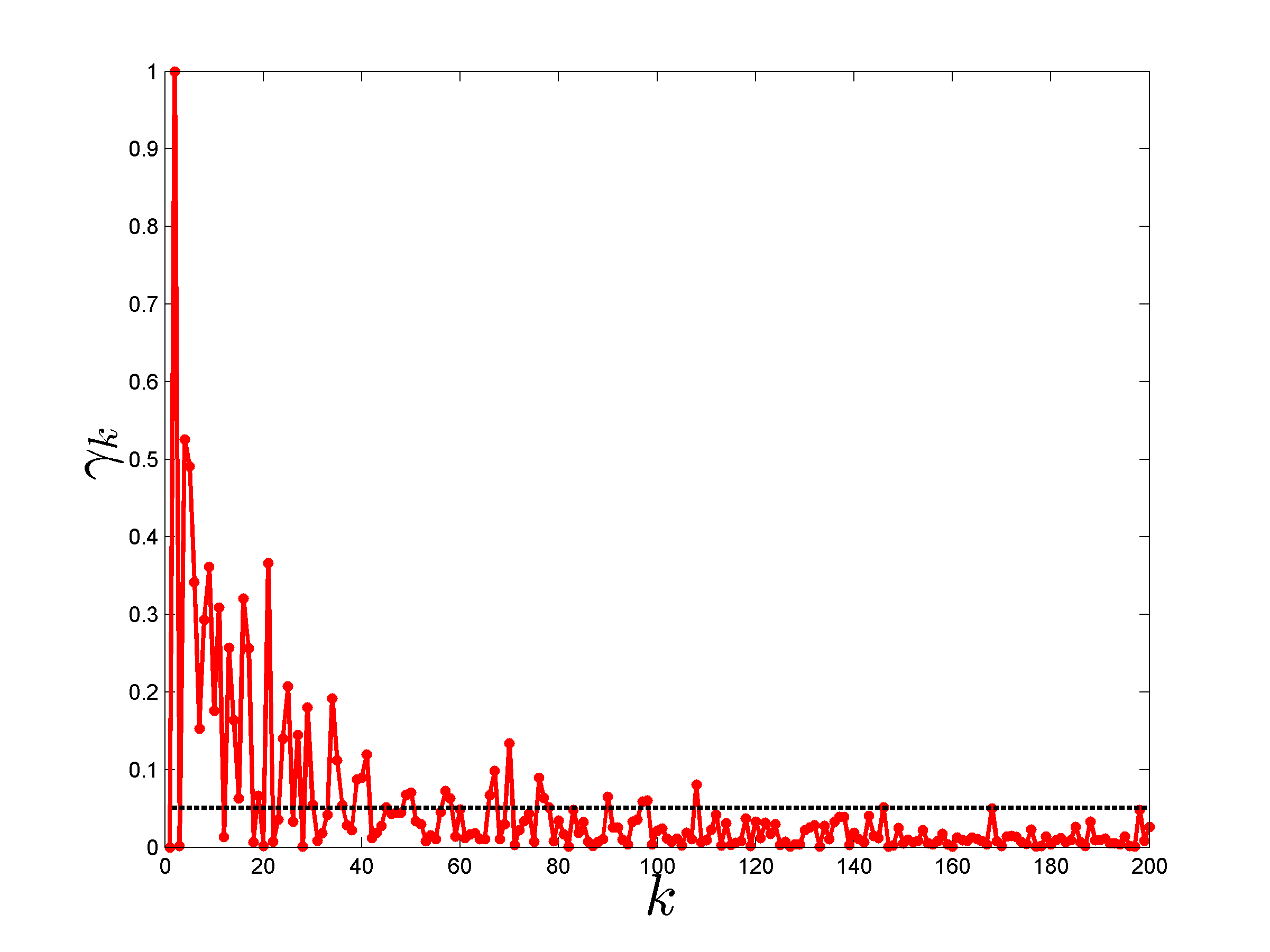}
\end{center}
\caption{Left: computational times of the greedy and lazy-greedy implementations; Right: effective size $\mg_k$ of $\SX_C$ as a function of $k$ (maximization of $\widehat I_{B,q}(\Zb_n)$ of Section~\ref{S:evaluation-cdf} with $\SX_Q=\SX_C=\Sb_{2\,048,d}$, $q=5$, $B=\sqrt{d}/2$; $d=5$ on the right panel).}
\label{F:greedy-lazyvsgreedy-IAB-2048-200}
\end{figure}

At each iteration of the greedy algorithm the maximum of $f(\Xb\cup\{\xb\})$ over all $\xb\in \SX_C\setminus\Xb$ must be found. As  presented in \cite{Minoux1977},  the submodularity of $f$ can be further exploited to restrict actual evaluation of $f$ to a proper subset of  $\SX_C\setminus\Xb$ (i.e., strictly included in $\SX_C\setminus\Xb$). Although the worst-case complexity of the modified algorithm is still $\SO(k\,C)$, important gains in computational complexity are observed in practice; see Figure~\ref{F:greedy-lazyvsgreedy-IAB-2048-200} for an illustration.

Denote by $\delta_\Xb(\xb) =f(\Xb\cup\{\xb\}) - f(\Xb) \geq 0$ the improvement of $f$ when $\xb$ is added to $\Xb$, and let $\Xb_n$ be the greedy solution at iteration $n$. Also,  remark that maximization of $f(\Xb_n\cup\{\xb\})$  in Step~3 of Algorithm~\ref{Algo:greedy} is equivalent to maximization of $\delta_{\Xb_n}(\xb)$.

By construction, $\Xb_i \subset \Xb_n$ for all $i<n$. Then, since $f$ is submodular, for all $i<n$, $\delta_{\Xb_n}(\xb) \leq \delta_{\Xb_i}(\xb)$; i.e., for each $\xb$ the increments $\delta_{\Xb_i}(\xb)$ decrease from iteration to iteration, as the size of $\Xb_i$ grows. At the first iteration, with $\Xb_0=\emptyset$, we compute all $\delta_{\Xb_0}(\xb)$ for all $\xb\in\SX_C$, establishing for each $\xb$ an upper bound $\overline{\delta}(\xb)$ on $\delta_{\Xb_k}(\xb)$ at subsequent iterations. These upper bounds are updated as follows.

Consider, at iteration $k$, scanning of the set $\SX_C\setminus\Xb_{k-1}$ to compute the solution of Step~3 of Algorithm~1. Let $\SL_{k-1}\subset\SX_C$ denote the set of points that are possible solutions of Step~3. Initialize $\SL_{k-1}=\SX_C\setminus\Xb_{k-1}$, and let $\xb_k^{\star\star}$ be its member with largest $\overline{\delta}(\xb)$.
While $\SL_{k-1} \neq\emptyset$, update $\overline{\delta}(\xb_k^{\star\star})=\delta_{\Xb_{k-1}}(\xb_k^{\star\star})$ and remove from $\SL_{k-1}$ all $\xb$ whose upper bound is smaller than or equal to $\overline{\delta}(\xb_k^{\star\star})$. When $\SL_{k-1}=\emptyset$, update $\Xb_{k-1}$ into $\Xb_k=\Xb_{k-1}\cup\{\xb_k^{\star\star}\}$.

This modified algorithm trades memory for computational power, requiring the storage of the most recently updated values of $\delta_\Xb(\xb)$ for all $\xb\in \SX_C\setminus\Xb$.

The left panel of Figure~\ref{F:greedy-lazyvsgreedy-IAB-2048-200} shows the computational times\footnote{All calculations are made with Matlab, on a PC with a clock speed of 3.3 GHz and 8 GB RAM, and operations are vectorized whenever possible.} of the greedy (solid lines) and lazy-greedy (dashed lines) maximizations\footnote{Note that, whenever  the solution at Step~3 is unique, the designs obtained in the two cases are identical.} of the non-decreasing submodular criterion $\widehat I_{B,q}$ that will be introduced in Section~\ref{S:evaluation-cdf}; see \eqref{empirical}. Notice the tremendous acceleration provided by the lazy-greedy implementation, contrasting with the fast linear increase of computational cost with $n$ for the greedy version. We can understand the lazy algorithm as inducing a decrease of the effective size of the candidate set $\SX_C$ at each iteration, from $C$ to $\mg\,C$, $\mg<1$.
Denote by $m_k$ the number of updates of $\xb_k^{\star\star}$ at iteration $k$ (with therefore $m_1=C$), and let $\mg_k=m_k/C$: the effective size of $\SX_C$ for the lazy-greedy algorithm is $m_k=C\,\mg_k$ at iteration $k$ and $\overline{\mg}_n\,C=(C/n)\,\sum_{k=1}^n \mg_k$ in $n$ iterations (i.e., for a $n$-point design). The right panel of Figure~\ref{F:greedy-lazyvsgreedy-IAB-2048-200} shows $\mg_k$ as a function of $k$ for the maximization of $\widehat I_{B,q}(\Zb_n)$, with an average (dashed horizontal line) of $\overline{\mg}_n \simeq 0.05$ for $n=200$. The behaviour observed is typical.

In the sequel, we shall always use the lazy-greedy algorithm to maximize a submodular set function.

\section{Covering measures}\label{S:covering}

\subsection{Definitions}

For any positive scalar $r$, we define the covering measure $\Phi_r(\Zb_n)$ of $\Zb_n$ by
$$
\Phi_r(\Zb_n)= \frac{\vol\{\SX\cap\left[\cup_{i=1}^n \SB(\zb_i,r)\right]\}}{\vol(\SX)} \,,
$$
with $\SB(\zb,r)$ the closed ball with center $\zb$ and radius $r$. For a given $\Zb_n$, consider also the function $r\in\mathds{R}^+ \ra F_{\Zb_n}(r)=\Phi_r(\Zb_n)$. $F_{\Zb_n}$ is non-decreasing and satisfies $F_{\Zb_n}(0)=0$ and $F_{\Zb_n}(r)=1$ for any $r\geq \CR(\Zb_n)$. If $X$ is distributed with the uniform probability measure $\mu$ on $\SX$, we have
\be\label{F_Xn}
\Pr\left\{X\in \cup_{i=1}^n \SB(\zb_i,r)\right\} = \Pr\{d(X,\Zb_n) \leq r\} = \int_{\{\xb\in\SX:\, d(\xb,\Zb_n)\leq r\}} \mu(\dd\xb) = F_{\Zb_n}(r) \,,
\ee
and $F_{\Zb_n}$ is the c.d.f.\ of the random variable $d(X,\Zb_n)$, supported on $[0,\CR(\Zb_n)]$. We will call $F_{\Zb_n}$ the distance c.d.f.\ and denote by $Q_\ma(\Zb_n)$ its $\ma$-quantile:
\bea
Q_\ma(\Zb_n) = \inf\{t: F_{\Zb_n}(t) \geq \ma\} \,,
\eea
with $Q_1(\Zb_n)=\CR(\Zb_n)$.
The set function $\Zb_n\subset\SX \ra \Phi_r(\Zb_n)$ is non-decreasing and satisfies $\Phi_r(\emptyset)=0$. Moreover, for any $\xb\in\SX$, the difference $\Phi_r(\Zb_n\cup\{\xb\})-\Phi_r(\Zb_n)$ is non-increasing with respect to $\Zb_n$, so that $\Phi_r$ is submodular.

In \cite{NoonanZ2020}, the authors also define design criteria based on the distribution of distances to design points, in particular the $(1-\mg)$-covering radius which corresponds to $Q_{1-\mg}(\Zb_n)$. Using approximations valid for large $d$ (see Remark~\ref{R:large-d-approximation}), they investigate the properties of $Q_{1-\mg}(\Zb_n)$ for different random designs, in particular designs formed by $n$ points i.i.d.\ in $\SC_d=[0,1]^d$ with a beta distribution. In \cite{NoonanZ2020b}, they consider full-factorial and $2^{d-1}$ fractional factorial designs, also for large $d$. We follow a different route, considering the values of $F_{\Zb_n}(r)$ taken at different $r$ and exploiting the submodularity of $\Phi_r$ to construct deterministic incremental designs $\Xb_n$ which, thanks to Theorem~\ref{th:NemhauserWF}, have guaranteed efficiency for all $n$ for the criterion considered. In contrast with \cite{NoonanZ2020}, our constructions are effective also for small $d$, as several examples will illustrate. In our numerical study, additionally to the covering radius, we will also assess performance through the covering $\ma$-quantile $Q_\ma$ for $\ma$ close to 1.

\subsection{A c.d.f.-based submodular covering criterion}\label{S:construction}

Denote by $f_{\Zb_n}$ the probability density function (p.d.f.) corresponding to $F_{\Zb_n}$. For any $B>0$, $q>-1$ and $\Zb_n\neq \emptyset$, define the integrated covering measure
\be
\IBq(\Zb_n) &=& \int_0^B r^q \, F_{\Zb_n}(r)\, \dd r \label{I} \\
&=&  \frac{1}{q+1} \left\{ B^{q+1} F_{\Zb_n}(B) - \int_0^B r^{q+1}\, f_{\Zb_n}(r)\, \dd r \right\} \,, \label{I2}
\ee
and set $\IBq(\emptyset)=0$.

The set function $\IBq:\, \Zb_n  \ra \IBq(\Zb_n)$ is non-decreasing and submodular, and satisfies $\IBq(\emptyset)=0$. As $\IBq$ satisfies the conditions of Theorem~1, its greedy maximization with Algorithm~1 provides sequences of incremental designs $\Xb_n$ with a $\IBq$-efficiency of at least 63.2\% for any $n$, see \eqref{*th1}.
Since $F_{\Zb_n}(B)=1$ for any $B\geq\CR(\Zb_n)$, maximizing $\IBq(\Zb_n)$ with respect to $\Zb_n$ for $B\geq\diam(\SX)$ is equivalent to minimizing $\int_0^B r^{q+1}\, f_{\Zb_n}(r)\, \dd r=\Ex_n\{R^{q+1}\}$ in \eqref{I2}, where the random variable $R$ has the p.d.f.\ $f_{\Zb_n}$ and where
\be\label{quant-error}
(\Ex_n\{R^{q+1}\})^{1/(q+1)} = E_{q+1}(\Zb_n)
\ee
is the $L^{q+1}$-mean quantization error induced by $\Zb_n$, see \cite{GrafL2000}.
It satisfies
$E_{q+1}(\Zb_n) \nearrow \CR(\Zb_n)$ as $q\ra\infty$. For $B$ and $q$ large enough, maximizing $\IBq(\Zb_n)$ should therefore provide designs with small values of $\CR(\Zb_n)$; moreover, the greedy maximization of $\IBq(\Zb_n)$ is equivalent to the greedy minimization of $E_{q+1}(\Zb_n)$, which is proved in \cite{LuschgyP2015} to ensure that  $E_{q+1}(\Zb_n)$ tends to zero at rate $n^{-1/d}$.
These observations motivate the investigations on the properties of $\IBq$ and its numerical implementation presented in the paper.

\subsection{Numerical implementation}\label{S:evaluation-cdf}

To evaluate $\IBq(\Zb_n)$ we substitute the empirical c.d.f.\ $\widehat F_{\Zb_n}$ for $F_{\Zb_n}(r)$ in \eqref{I}, where $\widehat F_{\Zb_n}$ is obtained by replacing $\mu$ in \eqref{F_Xn} by the uniform measure $\mu_Q$ supported on a finite subset $\SX_Q$ of $\SX$,
\bea
\SX_Q=\{\xb^{(1)}, \xb^{(2)},\ldots,\xb^{(Q)}\} \,.
\eea
The set $\SX_Q$ must be well spread over $\SX$; it may be a regular grid, or correspond to the first $Q$ points of a LDS in $\SX$.
The empirical distance c.d.f.\ $\widehat F_{\Zb_n}$ is based on the $Q$ distances $d_j(\Zb_n)=d(\xb^{(j)},\Zb_n)$, $j=1,\ldots,Q$. Denoting by $\overline{d}_{j}$ the truncated version of $d_{j}$, $\overline{d}_{j}(\Zb_n)=\min\{d_{j}(\Zb_n),B\}$, $j=1,\ldots,Q$, we obtain
\be\label{empirical}
\IBq(\Zb_n) \approx \widehat I_{B,q}(\Zb_n) = \frac{B^{q+1}}{q+1} - \frac{1}{Q(q+1)}\, \sum_{j=1}^{Q} \overline{d}_j^{q+1}(\Zb_n) \,.
\ee

\begin{remark}[A connection with clustering]\label{R:clustering}
When $B\geq\diam(\SX)$ ($B\geq \diam(\SX)/2$ when $\SX$ is convex and $\xb_1$ is its Chebyshev center), the greedy maximization of $\IBq(\Zb_n)$, $q>-1$, selects
\bea
\xb_{n+1} \in\Arg\max_{\xb\in\SX} \int_{\SC(\xb)} \left[ d^{q+1}(\zb,\Xb_n)-\|\xb-\zb\|^{q+1} \right] \,\mu(\dd\zb)
\eea
at iteration $n$, with $\SC(\xb)$ the cell with generator $\xb$ in the Voronoi partition of $\SX$ associated with $\Xb_n\cup\{\xb\}$.

One may notice that \eqref{empirical} implies that, when $B$ is large enough, the maximization of $\widehat I_{B,q}(\Zb_n)$ is equivalent to the center location problem \cite{Minoux1977} defined by the minimization of $\sum_i d_j^{q+1}(\Zb_n)$. It also corresponds to the minimization of the $L^{q+1}$-mean quantization error induced by $\Zb_n$, which, in this discrete setting where $\mu_Q$ is substituted for $\mu$, can be solved by clustering algorithms, see \cite{LekivetzJ2015}, including {\tt kmeans} and variants based of the Chebyshev centers of the Voronoi cells defined by the elements of $\Zb_n$ \cite{P-JSFdS2017}.
\fin
\end{remark}

\paragraph{Iterative update of the $d(\xb^{(j)},\Xb_n)$.}
Consider the greedy maximization of $\widehat I_{B,q}$. The criterion $\widehat I_{B,q}(\Xb_n \cup \{\xb\})$ optimized at step $n$ of the incremental construction involves the distances $d_j(\Xb_n\cup\{\xb^{(k)}\})$ for the current design $\Xb_n$ and all $\xb^{(k)}\in\SX_C$, the set of eligible design points. The $Q\times C$ matrix $\Db_n$ with elements $\{\Db_n\}_{j,k}=d_j(\Xb_n\cup\{\xb^{(k)}\})$, $\xb^{(k)}\in\SX_C$, can be computed recursively as follows. At initialization of Algorithm~1, $\{\Db_0\}_{j,k}$ is the $Q\times C$ matrix of inter-distances between the points of $\SX_Q$ and those of $\SX_C$. Then, when $\Xb_{n+1}=\Xb_n\cup\{\xb^{(k')}\}$ for some $\xb^{(k')}\in\SX_C$, each column $\{\Db_{n+1}\}_{\cdot,\ell}$ of $\Db_{n+1}$ is given by $\min\{\{\Db_n\}_{\cdot,\ell},\{\Db_0\}_{\cdot,k'}\}$, where the minimum is taken element-wise, for $\ell=1,\ldots,C$.
In practice, it is computationally more efficient to calculate $\Db_0^{(q+1)}$ and directly update $\Db_n^{(q+1)}$, with $\{\Db_n^{(q+1)}\}_{j,k}=[d_j(\Xb_n\cup\{\xb^{(k)}\})]^{q+1}$  for any $n\geq 0$.

The complexity of the evaluation of $\widehat I_{B,q}(\Xb_n \cup \{\xb^{(k)}\})$ for an $\xb^{(k)}\in\SX_C$ is of order $\SO(Q)$. The complexity of the greedy maximization of $\widehat I_{B,q}(\Zb_n)$ is thus of order $\SO((\kappa+n)\,CQ)$, where $\kappa\,CQ$ counts for the initial calculation of $\Db_0^{(q+1)}$; it becomes $\SO((\kappa+ n\,\overline{\mg}_n)\,CQ)$ for the lazy-greedy algorithm, with $\overline{\mg}_n \ll 1$, see Section~\ref{S:lazy-greedy}.

\begin{remark}[Large $d$ approximation]\label{R:large-d-approximation}
Using an Edgeworth expansion of the Central Limit Theorem, in \cite{ZhigljavskyN2020} the authors derive approximations of $F_{\xb}(r)$ that are valid for large $d$, with $F_\xb$ the distance c.d.f.\ for the one-point design $\{\xb\}$. For $X$ distributed with the uniform probability measure on $\SX$ and $\Zb_n$ a random design with $n$ i.i.d.\ elements $\zb_i$, we have $F_{\Zb_n}(r) = \Pr\{d(X,\Zb_n) \leq r\} = 1-\prod_{i=1}^n \left[1-F_{\zb_i}(r)\right]$; this is also exploited in \cite{NoonanZ2020}. However, the approximation of $F_\xb$ is not directly exploitable for the estimation of $\IBq(\Zb_n)$ \emph{for a given} $\Zb_n$ since in that case $F_{\Zb_n}(r)\neq 1-\prod_{i=1}^n \left[1-F_{\zb_i}(r)\right]$.
\fin
\end{remark}

\begin{remark}[Initialization]
When one is not interested in the performance of the smallest designs, with $n$ less than some $n_{\min}$, it seems preferable to directly choose the first $n_{\min}$ points by minimizing $\CR(\Zb_{n_{\min}})$, and initialize the greedy algorithm at this $n_{\min}$-point design. The batch optimization of $\CR(\Zb_{n_{\min}})$ is a difficult task, but several methods available are able to produce designs with reasonably good performance; see, e.g., \cite{P-JSFdS2017}. The investigation of properties and performance of incremental constructions initialized in this way is not considered in the present paper, where we consider the entire nested sequence, starting at $n=1$. This will be the subject of future studies, together with backwards constructions that start with a nearly optimal $n_{\max}$-point design and iteratively eliminate design points until an $n_{\min}$-points design is reached, for some $n_{\min}<n_{\max}$.
\fin
\end{remark}

\subsection{Parameter setting}\label{S:q&B}

\paragraph{Choice of $B$.}

Large enough values of $B$ have no influence on the maximization of $\IBq(\Zb_n)$. Indeed, take $B_2 \geq B_1 \geq \CR(\Zb_n)$; then $I_{B_2,q}(\Zb_n)=I_{B_1,q}(\Zb_n)+(B_2^{q+1}-B_1^{q+1})/(q+1)$ since $F_{\Zb_n}(r)=1$ for $B_1\leq r \leq B_2$. The simplest rule sets $B=\diam(\SX)$, guaranteeing that $F_{\Zb_n}(B)=1$ for any $\Zb_n$. A less simplistic rule is to choose $B$ of the order of magnitude of the largest possible value of $\CR(\Zb_n)$ over the range $[n_{\min},n_{\max}]$ of design sizes of interest, i.e., $\CR(\Zb_{n_{\min}})$.

When $\SX=\SC_d$ and $B\geq\sqrt{d}/2$, the first design point is necessarily taken at $\cb=(1/2,\ldots,1/2)$, the center of $\SC_d$. The unit cube $\SC_d$ can be covered by $m^d$ hypercubes with side $1/m$, and therefore by $m^d$ balls with radius $\sqrt{d}/(2m)$. Let $\CR_n^\star$ denote the minimum value of $\CR(\Zb_n)$ for designs of size $n$. Taking $m=\lfloor n^{1/d} \rfloor$, the largest integer smaller than or equal to $n^{1/d}$, we have $n\geq m^d$ and therefore the following upper bound holds:
\be\label{R^*}
\CR_n^\star\leq \CR_{m^d}^\star \leq  R^\star(n,d) = \frac{\sqrt{d}}{2\,\lfloor n^{1/d} \rfloor} \,.
\ee
This suggests taking $B=R^\star(n_{\min},d)$ instead of $B=\sqrt{d}/2$. The upper bound \eqref{R^*} is very pessimistic though, with $R^\star(1,d)=\sqrt{d}/2$ for all $d$ and the $k$th jump downwards of $R^\star(n,d)$ occurring at $n=(k+1)^d$. For example, the effect of choosing $B=R^\star(n_{\min},d)$ will only be effective for $n_{\min}\geq 32$ when $d=5$ and $n_{\min} \geq 1\,024$ when $d=10$. Our numerical studies indicate that it sometimes slightly improves the efficiency of the resulting designs when $n\geq 2^d$ (that is, for very large designs when $d$ is large). More interestingly, the first points selected are then no longer uniquely defined, which can be exploited to implement multi-start strategies.

\begin{figure}[ht!]
\begin{center}
 \includegraphics[width=.49\linewidth]{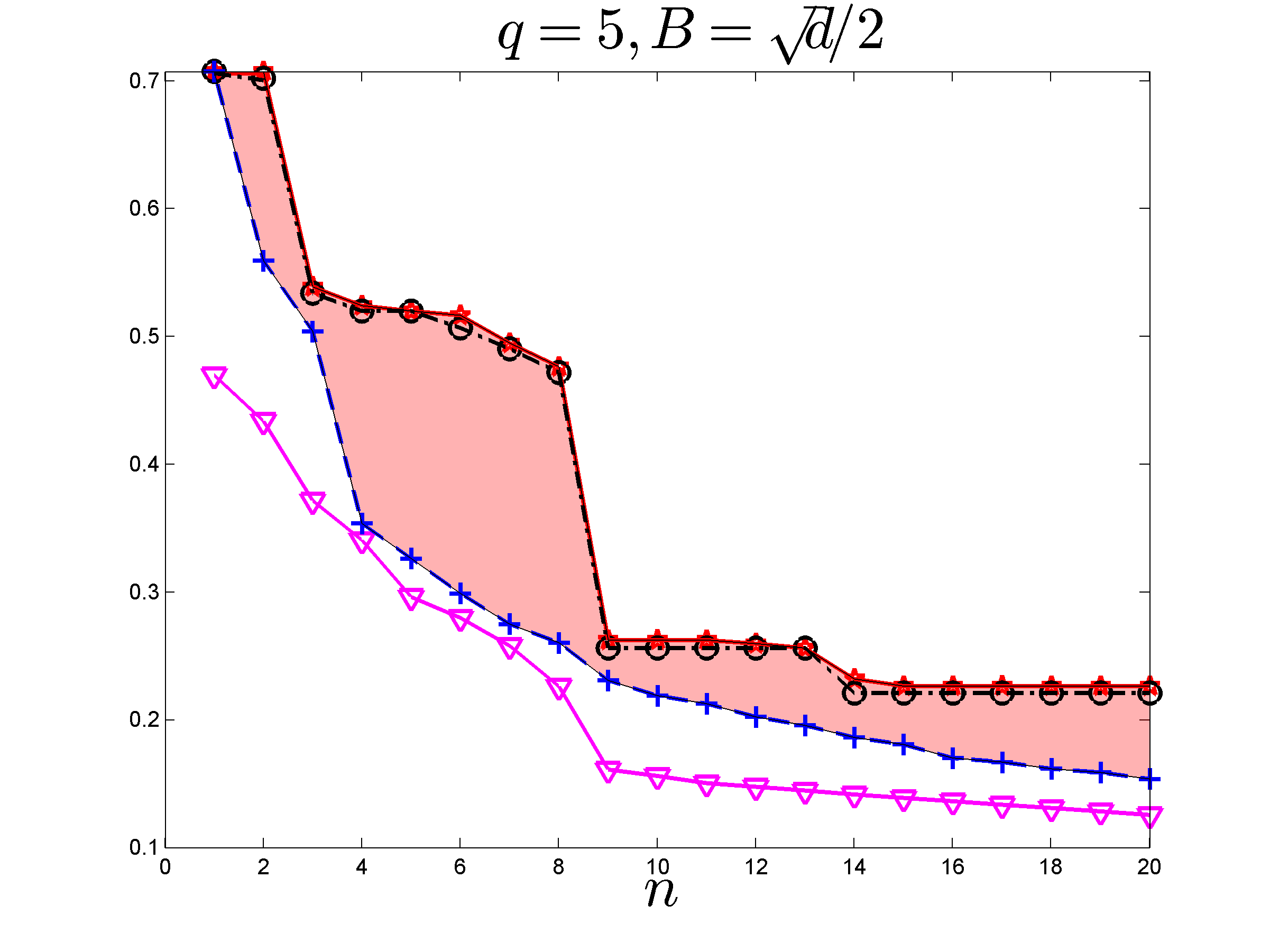} \includegraphics[width=.49\linewidth]{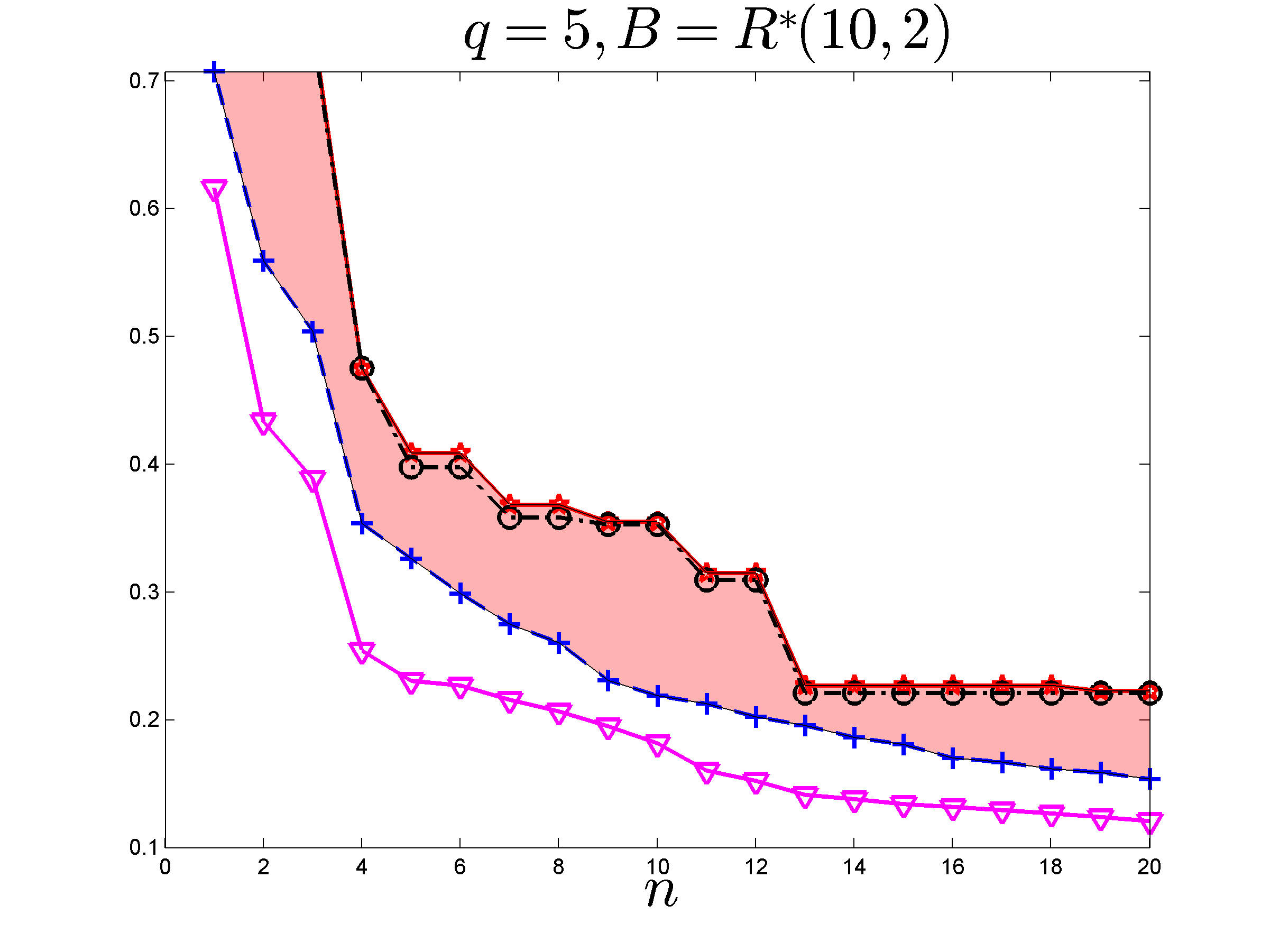}
\end{center}
\caption{$\CR(\Xb_n)$ (red $\bigstar$), $\CR_{[\SX_Q]}(\Xb_n)$ (black $\circ$), $\CR_n^\star$ (blue $+$), empirical value of $E_{q+1}(\Xb_n)$ \eqref{quant-error} (magenta $\triangledown$), for $\Xb_n$ obtained by greedy maximization of $\widehat I_{B,q}(\Xb_n)$.}
\label{F:CR_cdf_Q2048_d2}
\end{figure}

\begin{figure}[ht!]
\begin{center}
 \includegraphics[width=.49\linewidth]{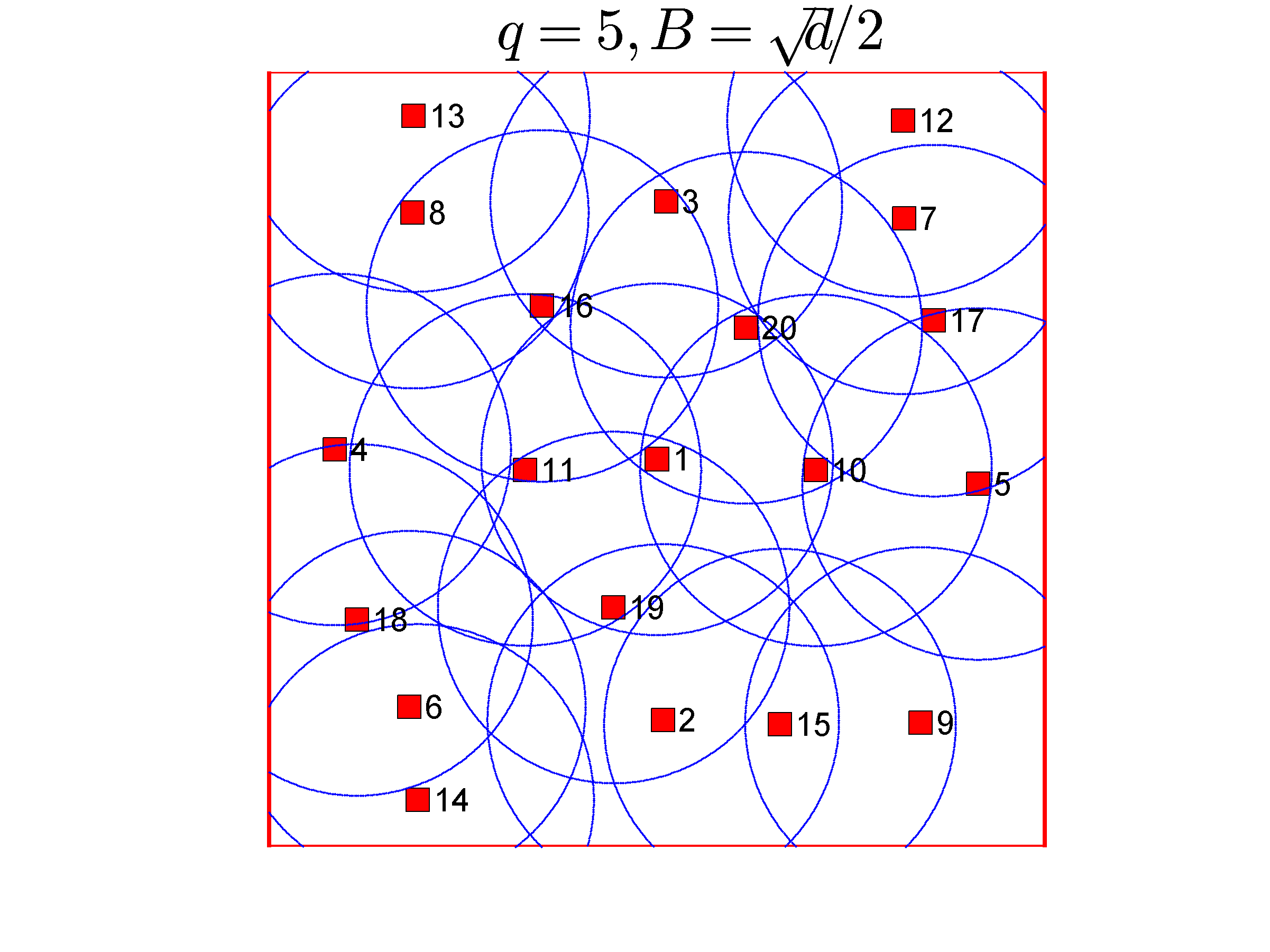} \includegraphics[width=.49\linewidth]{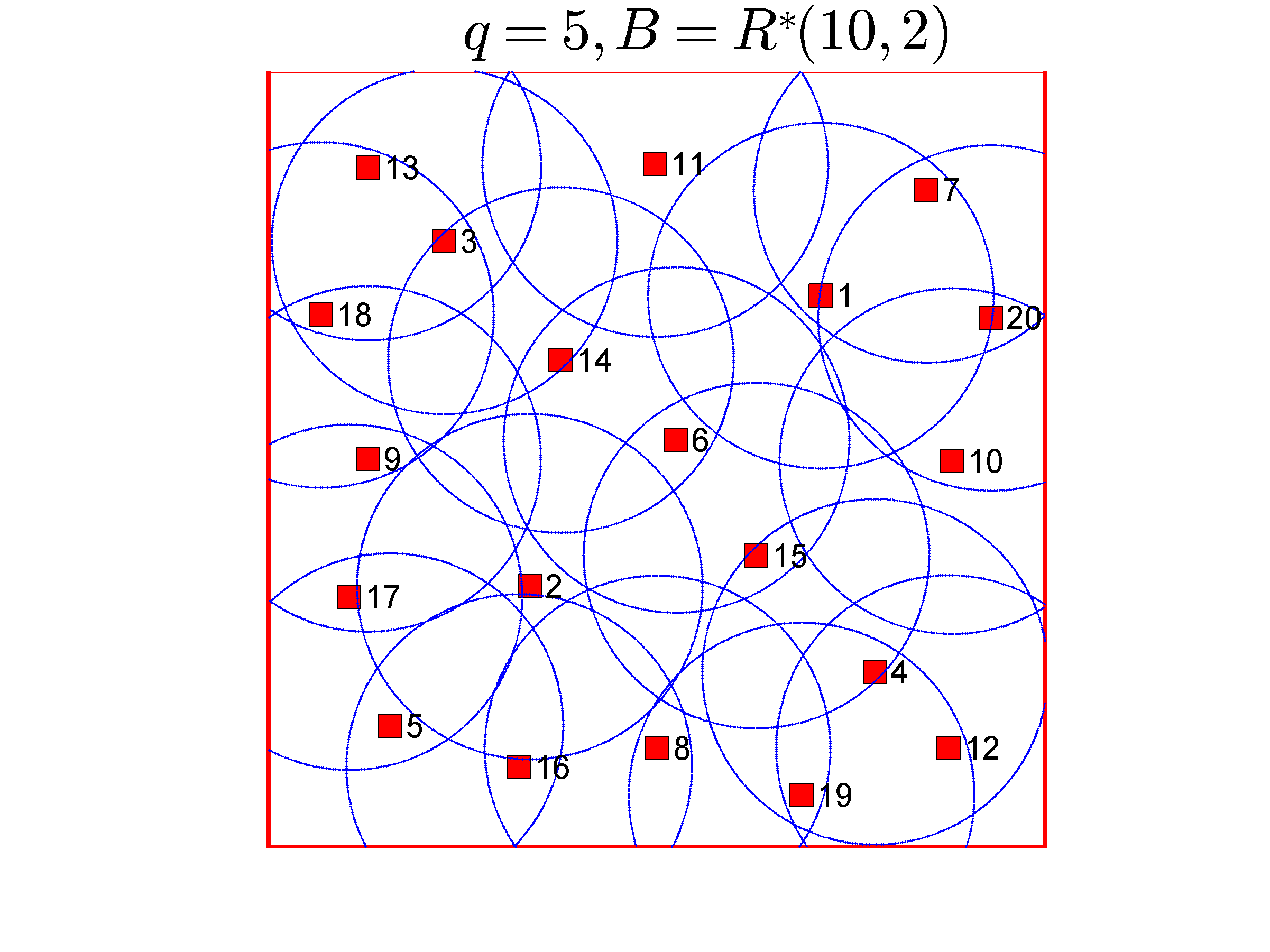}
\end{center}
\caption{$\Xb_{20}$ and circles centered at design points with radius $\CR(\Xb_{20})$; the order of selection of the points is indicated.}
\label{F:X20_20_Q2048_d2}
\end{figure}

Figures~\ref{F:CR_cdf_Q2048_d2} and \ref{F:X20_20_Q2048_d2} illustrate the impact of the choice of $B$ when $d=2$, $\SX_C=\SX_Q=\Sb_{2\,048,2}$ and $n_{\max}=20$. Figure~\ref{F:CR_cdf_Q2048_d2} presents the exact value of $\CR(\Xb_n)$ for $n$ between 1 and $n_{\max}$ (red solid curve with $\bigstar$) and its under approximation $\CR_{[\SX_Q]}(\Xb_n)$ given by \eqref{CR-approx} (black dashed curve with $\circ$). The blue dashed curve ($+$) presents the estimated\footnote{Exact for $n=1,2$, best (smallest) values obtained for each $n$ between 3 and 20 by running a {\tt kmeans}-type clustering algorithm on a $50\times 50$ regular grid in $\SX$; see \cite{P-JSFdS2017}. The values plotted are therefore not necessarily equal to the true values of $\CR_n^\star$, but we believe that the overestimation is negligible.} values of $\CR_n^\star$; the coloured area shows the region between $\CR_n^\star$ and $\CR(\Xb_n)$. The magenta curve ($\triangledown$) shows the empirical value of $E_{q+1}(\Xb_n)$ (a lower bound on $\CR_{[\SX_Q]}(\Xb_n)$, see \eqref{quant-error} in Section~\ref{S:construction}) for the uniform measure $\mu_Q$ on $\SX_Q$. Although the value $q=5$ is too small for $E_{q+1}(\Xb_n)$ to provide a good approximation of $\CR(\Xb_n)$, we can see that the greedy algorithm manages to ensure a reasonable decrease of $\CR(\Xb_n)$ along iterations. This decrease is more regular when $B=R^\star(n_{\min},d)$ than when $B=\sqrt{d}/2$. Figure~\ref{F:X20_20_Q2048_d2} shows the nested designs that are constructed and helps understanding the different behaviors of $\CR(\Xb_n)$: when $B=\sqrt{d}/2$ (left), $\xb_1$ corresponds to the optimal one-point design and is at the center of $\SC_2$, the next two points $\xb_2$ and $\xb_3$ are close to those for the optimal 3-point design $\{\xb_1,\widehat\xb_2,\widehat\xb_3\}$ with $\xb_1$ fixed at $(1/2,1/2)$; see the appendix. The situation is different when $B=R^\star(n_{\min},d)$ (right panel), and we see that sacrificing optimality at a given $n$ may be at the benefice of a more regular decrease of $\CR(\Xb_n)$: here the choice of the initial points $\xb_1$, $\xb_2$ and $\xb_3$ is much worse than with $B=\sqrt{d}/2$ (with $\CR(\Xb_1)=1.0054$ and $\CR(\Xb_2)=\CR(\Xb_3)=0.7442$), but the situation then improves, leading in particular to significantly smaller $\CR(\Xb_n)$ for $n\in\{5,6,7,8\}$.

\begin{remark}[Truncation of $B$ to $\CR(\Xb_n)$]\label{R:truncation_of_B}
For $B \geq \CR(\Xb_n)$, we have $I_{B,q}(\Xb_n\cup\{\xb\})=I_{\CR(\Xb_n),q}(\Xb_n\cup\{\xb\})+\{B^{q+1}-[\CR(\Xb_n)]^{q+1}\}/(q+1)$, where the second term does not depend on $\xb$. We may thus select $\xb_{n+1}$ as the maximizer of $I_{\min\{B,\CR(\Xb_n)\},q}(\Xb_n\cup\{\xb\})$.
When $q$ is large, the constant $C_n=\{B^{q+1}-[\CR(\Xb_n)]^{q+1}\}/(q+1)$ may become much larger than most terms in the sum \eqref{empirical}, so that ignoring it significantly improves numerical stability. Note that that the contribution of $C_n$ to $I_{B,q}(\Xb_n)$ must be suitably accounted for in the calculation of the bounds $\overline{\delta}(\xb^{(k)})$ on the increments in the lazy-greedy algorithm of Section~\ref{S:lazy-greedy}.
\fin
\end{remark}

\begin{remark}[A positive lower integration limit in $\IBq$]\label{R:b>0}
We discussed above the possibility of restricting the largest distances taken into account in $\IBq$. We can also think of restricting the contribution of the smallest distances by using a strictly positive lower integration limit $b$. The criterion is then more focussed on the influence of points of $\SX_Q$ at large distances from the design. Parameter $b$ must be taken larger than the smallest covering radius for the design sizes of interest, and we suggest to relate $b$ to a lower bound on $\CR_{n_{\max}}^\star$. Consider designs of size $n$ in $\SX=\SC_d$. Since the $n$ balls $\SB(\xb_i,\CR_n^\star)$ centered at the optimal design points cover $\SX$, $n V_d (\CR_n^\star)^d \geq \vol(\SX) = 1$, with $V_d$ the volume of the $d$-dimensional unit ball $\SB(\0b,1)$, $V_d = \pi^{d/2}/\Gamma(d/2+1)$.
This implies the following classical lower bound \cite[p.~150]{Niederreiter92}:
\be\label{R_*}
\CR_n^\star\geq R_\star(n,d)=(nV_d)^{-1/d} \,.
\ee
We may note that $V_d$ is maximum for $d=5$ (with $V_5 \simeq 5.2638$) and that $n^{1/d}\, R_\star(n,d)=\sqrt{d}/\sqrt{2\,\pi\,\e1}\, + \SO(1/\sqrt{d})$, $d\ra \infty$.
A reasonable choice is then $b=R_\star(n_{\max},d)$.
In practice, our numerical investigations indicate that this modification has marginal influence on performance, although it may slightly simplify the computations since the summation in \eqref{empirical} involves a smaller number of terms. It will not be used in the numerical experiments of Section~\ref{S:Examples}.
\fin
\end{remark}

\paragraph{Choice of $q$.}

\begin{figure}[ht!]
\begin{center}
 \includegraphics[width=.49\linewidth]{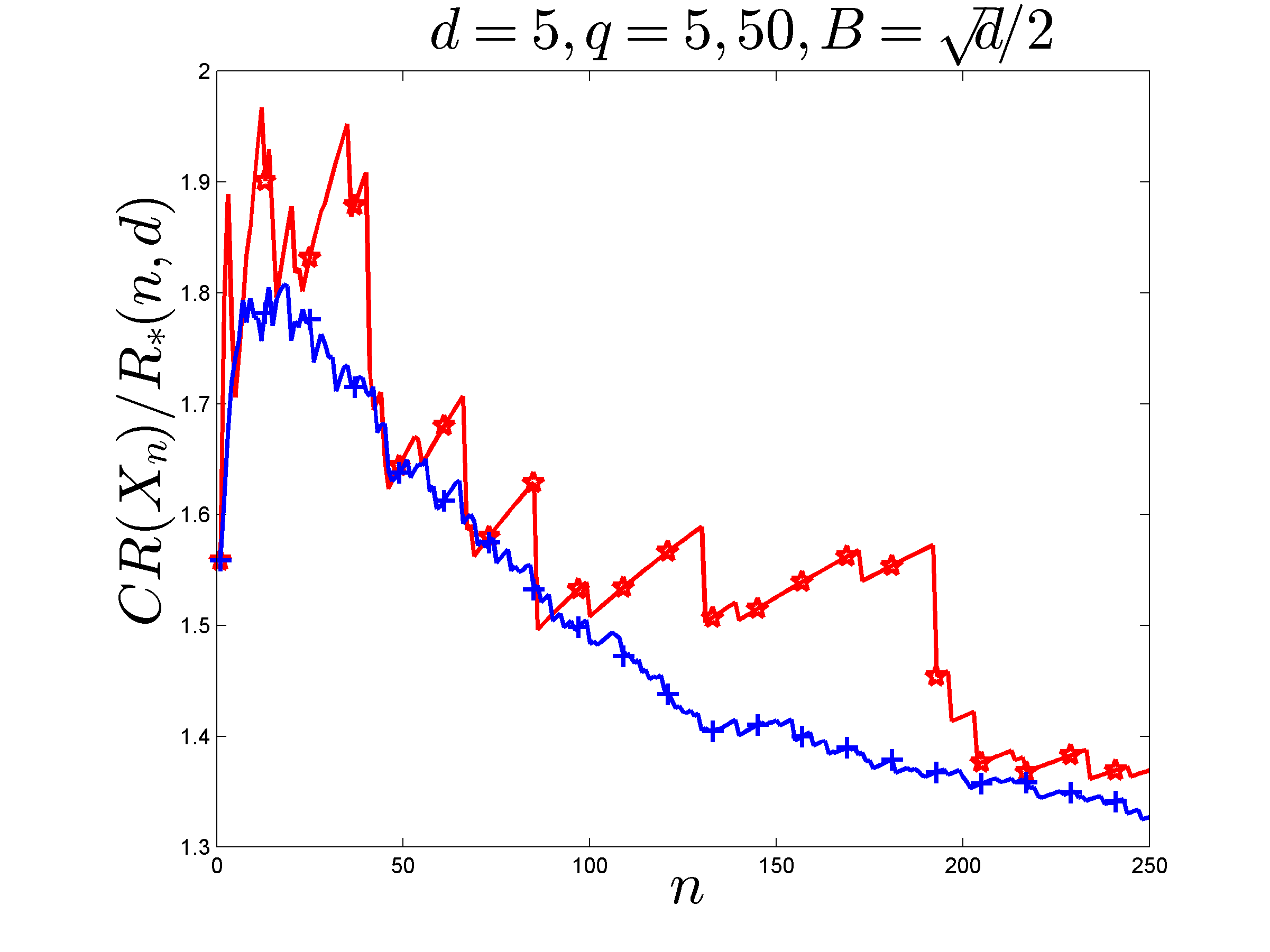} \includegraphics[width=.49\linewidth]{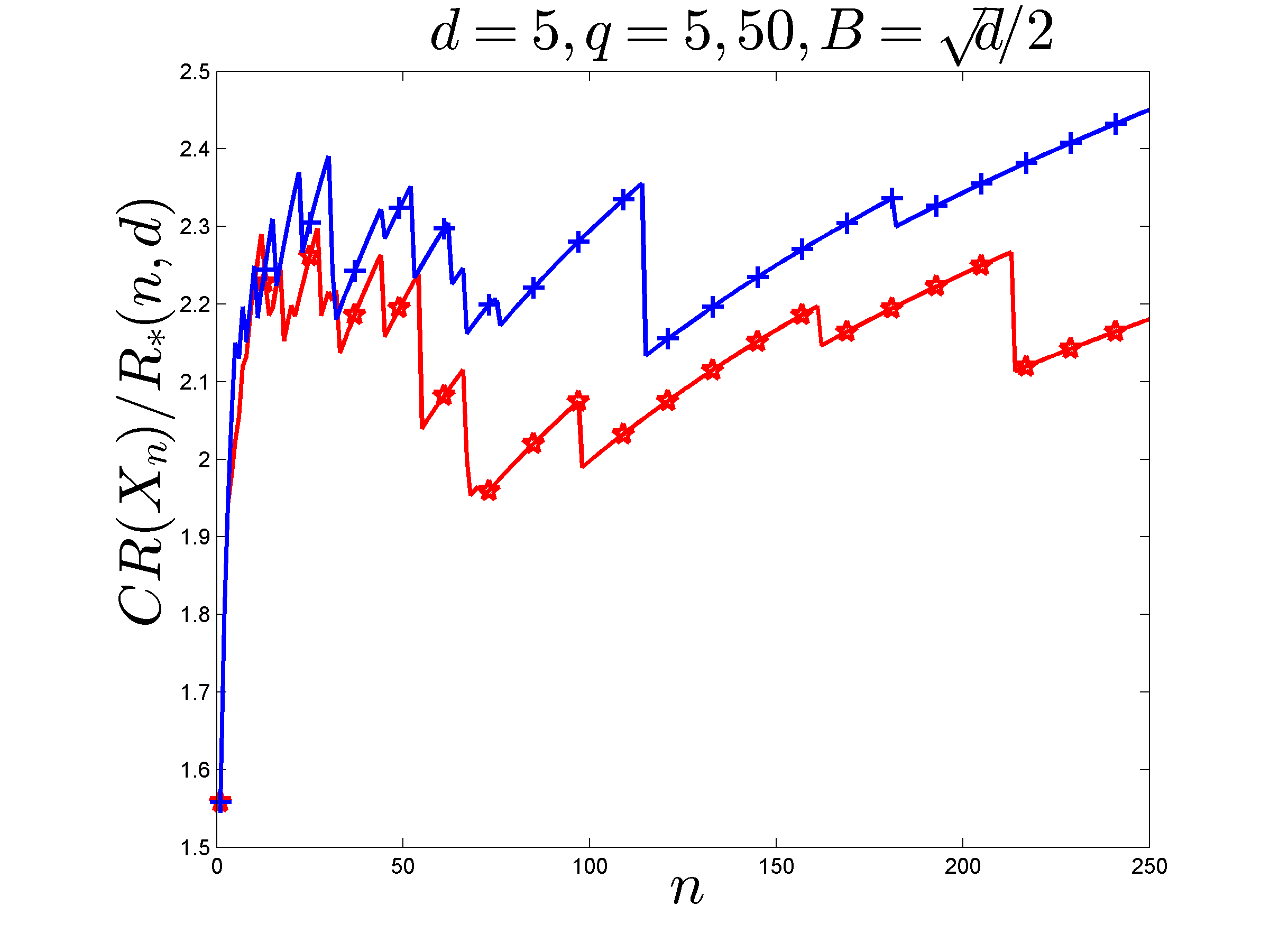}
\end{center}
\caption{Left: $\CR_{[\SX_Q]}(\Xb_n)/R_\star(n,d)$; Right: $\CR_{[\SX_N]}(\Xb_n)/R_\star(n,d)$, $n=1,\ldots,250$; $q=5$ (red $\bigstar$), $q=50$ (blue $+$).}
\label{F:d5_CQ2048_Bsqrt_q550_CR}
\end{figure}

The discussion in Section~\ref{S:construction} suggests that performance, as measured by the covering radius, should improve as $q$ increases. This is indeed the case for the criterion $\CR_{[\SX_Q]}(\Zb_n)$ defined by \eqref{CR-approx}. Whenever $B>\CR(\Zb_n)$, when $q$ tends to infinity a design $\Zb_n$ that maximizes  $\widehat I_{B,q}(\Zb_n)$ tends to minimize $\CR_{[\SX_Q]}(\Zb_n)$ as $\widehat I_{B,q}(\Zb_n)$ is focussed on the furthest points from $\Zb_n$ in $\SX_Q$, irrespectively of the presence of points in $\SX\setminus\SX_Q$ further away from $\Zb_n$. The connection with a clustering problem for the $L^{q+1}$-mean quantization error noticed in Remark~\ref{R:clustering} suggests that avoiding values of $q$ too large may lead to better generalization properties of the statistics on which $I_{B,q}$ is based when the number of data points on which it is evaluated is finite. Rather intensive numerical studies show that performance is robust with respect to the choice of $q$ and indicate that a value $q\in[5,25]$ is a convenient choice (see Figure~\ref{F:q_5-10-25} for an illustration). Figure~\ref{F:d5_CQ2048_Bsqrt_q550_CR} illustrates the impact of $q$, comparing the performances obtained for $q=5$ (red) and $q=50$ (blue) when $\SX=\SC_5$ ($d=5$), with $\SX_C=\SX_Q=\Sb_{2\,048,d}$ and $B=\sqrt{d}/2$. We normalize $\CR(\Xb_n)$ by dividing it by the lower bound \eqref{R_*} on the optimal (minimum) value $\CR_n^\star$. On the left panel, $\CR(\Xb_n)$ is approximated by $\CR_{[\SX_Q]}(\Xb_n)$, on the right panel it is approximated by $\CR_{[\SX_N]}(\Xb_n)$, with $\SX_N$ given by $2^{18}$ points of a scrambled Sobol' sequence complemented by a $2^d$ full factorial design (which gives $N=262\,176$). The choice $q=50$ is preferable to $q=5$ in terms of $\CR_{[\SX_Q]}(\Xb_n)$, but is significantly worse for $\CR_{[\SX_N]}(\Xb_n)$, which is of course a better approximation of $\CR(\Xb_n)$.

\paragraph{Choices of $\SX_C$ and $\SX_Q$.}
The problem considered may impose constraints on the set $\SX_C$ of eligible points. For instance, in Section~\ref{S:Examples} we force the algorithm to choose points within a Latin hypercube design, and use $\IBq$ to define an order in this design. When no such constraints are enforced, any finite set evenly distributed on $\SX$ can be used and in the paper we most often use points from a LDS.

A set of points evenly distributed on $\SX$ is required for $\SX_Q$, with $Q$ large being more crucial than having $C$ large. In particular, $\SX_Q$ must sample the regions expected to be the furthest away from designs with good $\CR$. For that reason, when $\SX=\SC_d$, $\SX_Q$ should contain the vertices of $\SC_d$. Given the factor $CQ$ in the algorithmic complexity of the construction and the size $Q\times C$ of the matrices involved, see Section~\ref{S:evaluation-cdf}, computational constraints impose an upper bound on the product $C\,Q$. Reducing the size of $\SX_C$ to a minimal but acceptable size then allows to use larger sets for $\SX_Q$, which is beneficial for the quality of the approximation of $\IBq(\Zb_n)$ by $\widehat I_{B,q}(\Zb_n)$. Our numerical experiments suggest using $\SX_C\subset\SX_Q$ with the rule of thumb $Q=2\,C$.

\section{Spacings and coffee-house design}\label{S:CH}

\paragraph{Coffee-house design.}
The algorithmic construction presented in the previous section is based on the distance c.d.f.\ $F_{\Zb_n}$.
In this section we present a second family of incremental algorithms, directly related to the most common geometrical criteria, the covering and packing radii. Even if these criteria  are not submodular, and thus Theorem~\ref{th:NemhauserWF} cannot be invoked to guarantee a lower bound on the efficiency of their greedy optimization, an efficiency of 50\% can still be guaranteed.

Coffee-house designs \cite{Muller2000}, \cite[Chap.~4]{Muller2007} greedily maximize $\PR(\Zb_n)$ for $n>1$; see \cite{KennardS69} for an early suggestion. When $\SX$ is convex, the first design point $\xb_1$ is usually chosen at the Chebyshev center of $\SX$ (the center of the minimal-radius ball enclosing $\SX$); then, at any $n\geq 1$, $\xb_{n+1}\in\Arg\max_{\xb\in\SX} d(\xb,\Xb_n)$. Denote by $\PR_n^\star$ the optimal (maximum) value of $\PR(\Zb_n)$ for a $n$-point design in $\SX$. The two ratios $\CR_n^\star/\CR(\Zb_n)$ and $\PR(\Zb_n)/\PR_n^\star$ are less than one by construction and quantify the efficiency of a design $\Zb_n$ in terms of its covering and packing radii.
It is remarkable that the simple greedy coffee-house construction ensures the following property (irrespectively of the choice made for $\xb_1$):
\be\label{efficiency-CH}
\frac12 \leq \frac{\CR_n^\star}{\CR(\Xb_n)} \leq 1 \ (n\geq 1) \ \  \mbox{ and } \ \ \frac12 \leq \frac{\PR(\Xb_n)}{\PR_n^\star}\leq 1 \ (n\geq 2) \,.
\ee
Indeed, by construction $\PR(\Xb_{n+1})=d(\xb_{n+1},\Xb_n)/2=\CR(\Xb_n)/2$ for all $n\geq 1$.
Take any $n$-point design $\Zb_n=\{\zb_1,\ldots,\zb_n\}$ in $\SX$. From the pigeonhole principle, one of the balls $\SB(\zb_i,\CR(\Zb_n))$ must contain two points $\xb_i,\xb_j$ of $\Xb_{n+1}$, implying that $\PR(\Xb_{n+1}) \leq \|\xb_i-\xb_j\|/2 \leq \CR(\Zb_n)$. Therefore, $\CR_n^\star \geq \PR(\Xb_{n+1})=\CR(\Xb_n)/2$. Similarly, $\PR_{n+1}^\star \leq  \CR_n^\star \leq \CR(\Xb_n) =2\,\PR(\Xb_{n+1})$. The original proof is given in \cite{Gonzalez85}.

As for the maximization of $\IBq(\Zb_n)$ in Section~\ref{S:covering}, the implementation of Algorithm~1 is much facilitated when
$\xb_{n+1}\in\Arg\max_{\xb\in\SX_C} d(\xb,\Xb_n)$ with $\SX_C$ a finite set of candidates. The efficiencies given in \eqref{efficiency-CH} remain valid provided that $\CR(\Xb_n)$ is approximated by $\CR_{[\SX_C]}(\Xb_n)$ and that $\CR_n^\star$ and $\PR_n^\star$ are relative to optimal $n$-point designs in $\SX_C$. Contrary to the maximization of $\widehat I_{B,q}(\Zb_n)$ in Section~\ref{S:evaluation-cdf}, the greedy maximization of $\PR(\Zb_n)$ does not require computing a $Q\times C$ matrix of inter-distances, but only the update of the $C$ distances $d(\xb^{(i)},\Zb_n)$ for the $\xb^{(i)}$ in $\SX_C$. For that reason, the size $C$ of the candidate set can be taken much larger than in Section~\ref{S:q&B}.

Denote by $\rho(\Zb_n)$ the mesh-ratio of $\Zb_n$, defined by
\bea
\rho(\Zb_n)= \frac{\CR(\Zb_n)}{\PR(\Zb_n)} \ (n\geq 2)\,.
\eea
If a sequence of $n$-points designs $\Zb_n$ is such that $\rho(\Zb_n)$ is uniformly bounded, then the sequence is said to be quasi-uniform; see, e.g., \cite{BondarenkoHS2014}.
The inverse of $\rho(\Zb_n)$ is sometimes called uniformity measure \cite{DeMarchi2003}.
When $\SX$ is convex it cannot be covered by two or more non-overlapping balls having their centers in $\SX$, and $\PR(\Zb_n)<\CR(\Zb_n)$, implying that $\rho(\Zb_n)>1$. When $\Xb_n$ is a coffee-house design, $\PR(\Xb_{n+1})=\CR(\Xb_n)/2$ for $n\geq 1$, and since $\CR(\Zb_n)\geq\CR(\Zb_{n+1})$ for all $n$ and any nested designs $\Zb_n\subset\Zb_{n+1}$, $\rho(\Xb_n)$ satisfies
\bea 
1 \leq \rho(\Xb_n)\leq 2 \ (n\geq 2) \,.
\eea
That $\rho(\Xb_n)$ is upper-bounded by 2 is a very strong guarantee of coffee-house design, which does not hold for our c.d.f.-based construction. This powerful property is certainly related to the bounded optimality gaps for both the covering and packing radii. The method presented in the previous section, tailored to control $\CR$ only, is unable to guarantee a large spacing between design points, which can translate into large values\footnote{Values larger than 3 have been observed in Example~1 of Section~\ref{S:Examples}.} of $\rho(\Xb_n)$.

\paragraph{Spacings and boundary-phobic coffee-house design.}
By construction, a coffee-house design places design points on the boundary $\partial\SX$ of $\SX$, in contradiction with the objective of constructing designs with low values of $\CR(\Zb_n)$. For that reason, we consider below a modified method that forces design points to stay away from $\partial\SX$.

\begin{figure}[ht!]
\begin{center}
 \includegraphics[width=.75\linewidth]{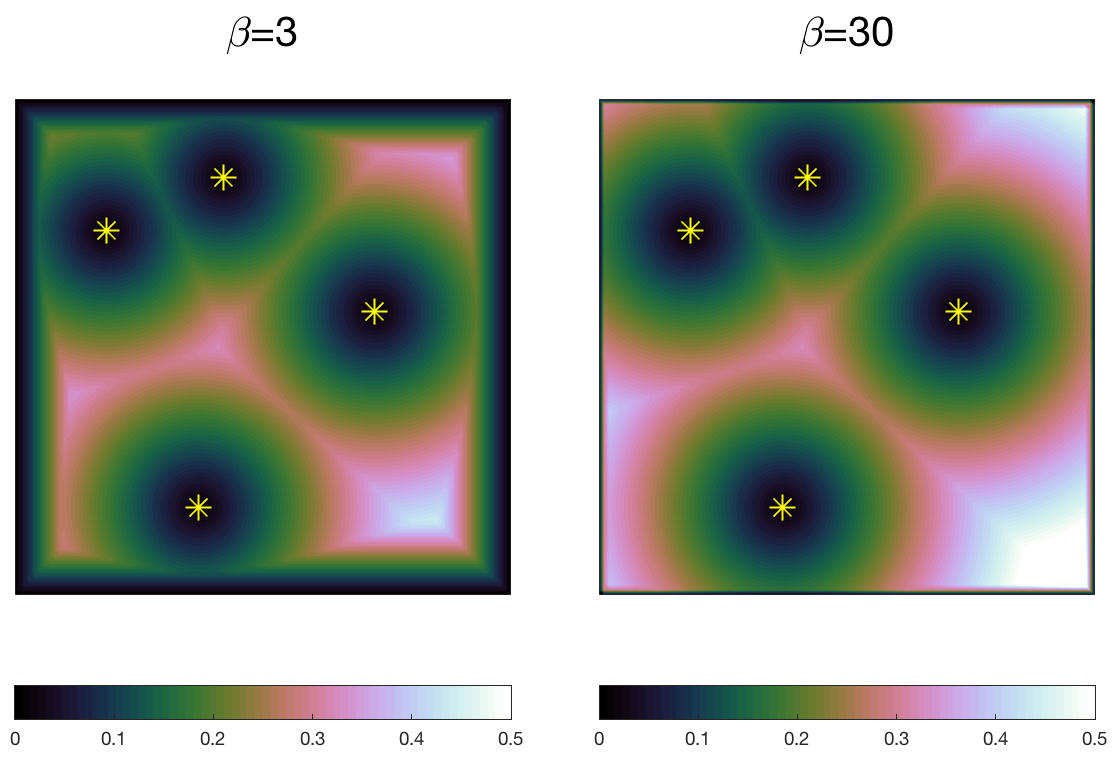}
\end{center}
\caption{Boundary-phobic compromise distance $D_\beta(\xb,\Zb_n)$ for a 4-point design (yellow stars) in $\SX=[0,1]^2$.}
\label{F:D_beta}
\end{figure}

Following \cite{Janson87}\footnote{In \cite{Janson87}, $\SX$ is only assumed to be bounded and bounded convex sets other than $\SB(\0b,1)$ are also considered.}, we define the maximal spacing $S(\Zb_n)$ as the radius of the largest ball contained in $\SX$ and not intersecting $\Zb_n$,
\bea
S(\Zb_n) = \sup \left\{r: \exists \xb \mbox{ such that } \xb \oplus r\,\SB(\0b,1) \subset \SX \setminus \Zb_n\right\}\,,
\eea
with $\oplus$ denoting the Minkowski sum (for two subsets $\SA_1$ and $\SA_2$ of $\mathds{R}^d$, $\SA_1\oplus\SA_2=\{\xb_1+\xb_2:\, \xb_1\in\SA_1,\, \xb_2\in\SA_2\}$.)
We slightly extend this notion by introducing a parameter $\beta$ that controls the ratio between the distance to the design and the distance\footnote{When $\SX=\SC_d$, an anonymous referee suggested to penalize differently distances to the vertices and distances to the boundary of $\SC_d$ (or more generally to the $d'$-dimensional faces of $\SC_d$, $1<d'<d$), using different metrics; we have not explored this possibility here.} to $\partial\SX$, and define the $\beta$-spacing of $\Zb_n$, for $\beta>0$, as
\bea
S_\beta(\Zb_n) = \sup \left\{r: \exists \xb\in\SX \mbox{ such that } d(\xb,\Zb_n) \geq r \mbox{ and } d(\xb,\partial\SX) \geq \frac{r}{\beta} \right\}
= \sup_{\xb\in\SX} D_\beta(\xb,\Zb_n) \,,
\eea
where
\be\label{Dbeta}
D_\beta(\xb,\Zb_n) = \min\left\{d(\xb,\Zb_n)\,, \ \beta\, d(\xb,\partial\SX)\right\} \,, \ \xb\in\SX\,,
\ee
and $d(\xb,\partial\SX)=\inf_{\zb\in\partial\SX} \|\xb-\zb\|$. Figure~\ref{F:D_beta} shows the variations of $D_\beta(\xb,\Zb_n)$ for a 4-point design $\Zb_n$ in the square $[0,1]^2$ for two different values of $\beta$. We have $S_1(\Zb_n)=S(\Zb_n)$ and define $S_\infty(\Zb_n)=\CR(\Zb_n)$.
We also define
\be\label{P-beta}
P_\beta(\Zb_n)= \min_{\zb_i\neq\zb_j\in\Zb_n} \frac12\, \min\left\{ \|\zb_i-\zb_j\|\,, \ \beta\, d(\zb_i,\partial\SX) \right\} \mbox{ and }
\rho_\beta(\Zb_n)=\frac{S_\beta(\Zb_n)}{P_\beta(\Zb_n)} \ (n\geq 2)\,,
\ee
and set $P_\infty(\Zb_n)=\PR(\Zb_n)$, so that $\rho_\infty(\Zb_n)=\rho(\Zb_n)$.

The coffee-house algorithm can be straightforwardly extended to the greedy maximization of $P_\beta(\Zb_n)$, $\beta>0$, using $\xb_{n+1}\in\Arg\max_{\xb\in\SX} D_\beta(\xb,\Xb_n)$ for any $n\geq 1$ (and if we use the convention $\Xb_0=\emptyset$ and $d(\xb,\emptyset)=+\infty$ for any $\xb$, the same rule sets $\xb_1$ at the Chebyshev center of $\SX$ when $\SX$ is convex and to a point of the medial axis \cite{Blum67} at maximum distance from $\partial\SX$ in general).
Following the same steps as for coffee-house design, we obtain the following property, whose proof is given in the appendix.

\begin{theorem}\label{TH:CH} Let $\SX$ be a compact and convex subset of $\mathds{R}^d$ and let $\Xb_n$ be a design obtained with the following greedy construction: $\Xb_0=\emptyset$ and $\xb_{k+1} \in\Arg\max_{\xb\in\SX} D_\beta(\xb,\Xb_k)$, $k=0,1,\ldots,n-1$, where $D_\beta$ is defined in \eqref{Dbeta}, $\beta>0$, and $d(\xb,\emptyset)=+\infty$. Then $P_\beta(\Xb_{n+1}) = (1/2)\,S_\beta(\Xb_n)$ for $n\geq 1$ and
\be\label{SalphaPalphamainineq}
\frac12 \leq \frac{S_{\beta,n}^\star}{S_\beta(\Xb_n)} \leq 1 \ (n\geq 1) \ \  \mbox{ and } \ \ \frac12 \leq \frac{P_\beta(\Xb_n)}{P_{\beta,n}^\star}\leq 1 \ \,, \ 1 \leq \rho_\beta(\Xb_n) \leq 2 \
(n\geq 2) \,,
\ee
where $S_{\beta,n}^\star=\min_{\Zb_n\subset\SX} S_\beta(\Zb_n)$ and $P_{\beta,n}^\star=\max_{\Zb_n\subset\SX} P_\beta(\Zb_n)$.
\end{theorem}


Algorithm~2 of \cite{ShangA2020}, also involving the domain boundary, corresponds to the greedy maximization of $P_\beta(\Zb_n)$ in $\SX=\SC_d$, for the value $\beta=2\,\sqrt{2\,d}$ (chosen in \cite{ShangA2020} by trial and error). Considering that the covering radius is of special importance, we recommend a different choice for $\beta$ when $\SX=\SC_d$. Since $\xb_1=\1b_d/2$, the center of $\SX$, when $\beta=\infty$ the second point $\xb_2$ coincides with a vertex of $\SX$. We suggest to choose $\beta$ depending on the maximum target design size $n_{\max}$, such that $\xb_2$ be at distance $R_\star(n_{\max},d)$ from a vertex of $\SX$, with $R_\star(n_{\max},d)$ the lower bound on $\CR_{n_{\max}}^\star$ given by \eqref{R_*}.
This implies $d(\xb_2,\partial\SX)=R_\star(n_{\max},d)/\sqrt{d} = \|\xb_2-\xb_1\|/\beta$ and gives
\be\label{beta*}
\beta = \beta_\star(n_{\max},d)=\frac{d}{2\,R_\star(n_{\max},d)} - \sqrt{d} \,.
\ee
The left panel of Figure~\ref{F:CR_bound_Faure_Sobol} in the appendix shows $\beta_\star(n,d)$ as a function of $d$ for $n=50$ (red $\bigstar$), $n=100$ (blue $\triangledown$) and $n=200$ (black $\circ$); the curve with magenta $\times$ corresponds to the values suggested in \cite{ShangA2020}. We see that as $n$ grows $\beta_\star(n,d)$ increases (since we expect designs to better cover $\SX$), and as $d$ becomes large the value of $\beta_\star(n,d)$ exhibits a very slow growth after an initial fast decay (being almost constant for large values of $n$).

\section{Numerical study} \label{S:Examples}

This section assesses the performance of the design algorithms presented in the previous sections.
Designs are compared in terms of the covering radius $\CR(\Xb_n)$ and of the covering $\ma$-quantile $Q_\ma(\Xb_n)$, with $\ma=0.99$; we will show plots of the evolution of these criteria as the greedy constructions unfold.
We stress that for the incremental design methods studied in this paper, the entire evolution of the performance indicators over a target range of design sizes is of interest, and not just their  value for a given final design size. Since $\CR(\Xb_n)$ and $Q_\ma(\Xb_n)$ ideally decrease as $n^{-1/d}$ all our plots will show normalized versions,
$\CR(\Xb_n)/R_\star(n,d)$, see \eqref{R_*}, and $n^{1/d}\,Q_\ma(\Xb_n)$. An ideal design method should lead to trajectories close to an horizontal line at a small value.

We compare the performance of designs $\Xb_n^\star$ that incrementally maximize $\widehat I_{B,q}(\Xb_n)$ to three families of constructions: (\textit{i}) prefixes of low discrepancy Halton ($\Hb_{n,d}$) and Sobol' ($\Sb_{n,d}$) sequences; (\textit{ii}) $\Xb_n^{CH,\beta}$ obtained by greedy maximization of $P_\beta(\Zb_n)$ given by \eqref{P-beta}, with $\beta \in \{\beta_\star(n_{\max},d),\, 2\sqrt{2},\, \infty\}$; and (\textit{iii}) two incremental constructions (designs $\Xb_n^{VD}$ and $\Xb_n^{RD}$) based on a direct relaxation of the covering radius recently proposed in \cite{PZ2019-JSC}, see precise definitions in the appendix.

Two case-studies are considered. In Example~1 the design space has a simple geometry embedded in a relatively  large dimensional space, $\SX = \SC_{10}=[0,1]^{10}$, while
Example~2 considers a more challenging topology involving a non-convex (annular) domain $\SX\subset\mathds{R}^2$. In both examples, the parameters of $\widehat I_{B,q}(\Xb_n)$ are $B=\diam(\SX)$ and $q=10$ (we also use $q=10$ for $\Xb_n^{VD}$ and $\Xb_n^{RD}$); design sizes exceed the range prescribed by the $n=10\,d$ rule of \cite{LoeppkySW2009}\footnote{In Example~1, very small designs with $n\leq 10$ are not meaningful and are excluded from the analysis.}.

\paragraph{Example~1 ($\SX=\SC_{10}$).}
In this example, the sets $\SX_C$ and $\SX_Q$ are always of the form $\SX_C=\Sb_{C,d}$ and $\SX_Q=\Sb_{Q',d}\cup\{0,1\}^d$ (for $\Xb_n^{VD}$ and $\Xb_n^{RD}$ a scrambled Sobol' sequence is used for $\SX_C$, as their constructions require that $\SX_C\cap\SX_Q=\emptyset$).
We use $C=8\,192$ and $Q'=16\,384$ for all methods. Note that the size of the candidate set for the construction of $\Xb_n^{CH,\beta}$ exceeds the one suggested in \cite{ShangA2020} ($C=1\,000\,d+2\,n_{\max}$). The value of $\CR(\Zb_n)$ is approximated by $\CR_{[\SX_N]}(\Zb_n)$ given by \eqref{CR-approx}, with $\SX_N$ consisting of $2^{18}$ points of a scrambled Sobol' sequence complemented by the $2^d$ full factorial design $\{0,1\}^d$ ($N= 263\,168$), and $Q_\ma(\Xb_n)$ is approximated by its empirical estimate based on $\SX_N$. All plots are normalized and consider $n\in[10,200]$.

It is noticeable that for all designs constructed in this example $\CR_{[\SX_N]}(\Xb_n)=\CR_{\{0,1\}^d}(\Xb_n)$, indicating that only the vertices of $\SC_d$ matter when evaluating $\CR$ for these designs. This dominant effect of the vertices on the values of $\CR(\Xb_n)$ for large $d$ and $n\ll 2^d$ flags the limitation of the covering radius as a measure of the global space-fillingness of designs in the hypercube.

\begin{figure}[ht!]
\begin{center}
 \includegraphics[width=.49\linewidth]{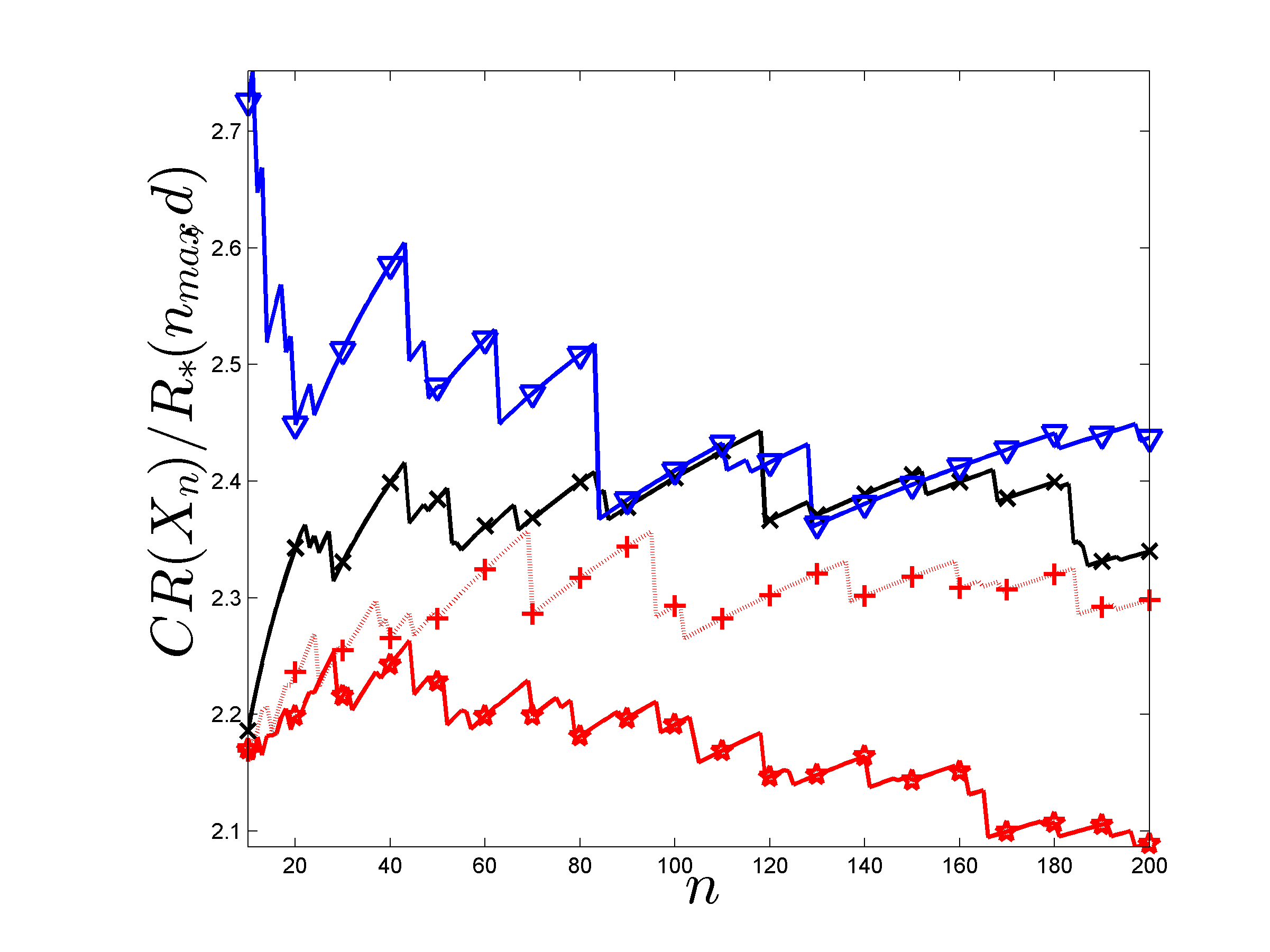} \includegraphics[width=.49\linewidth]{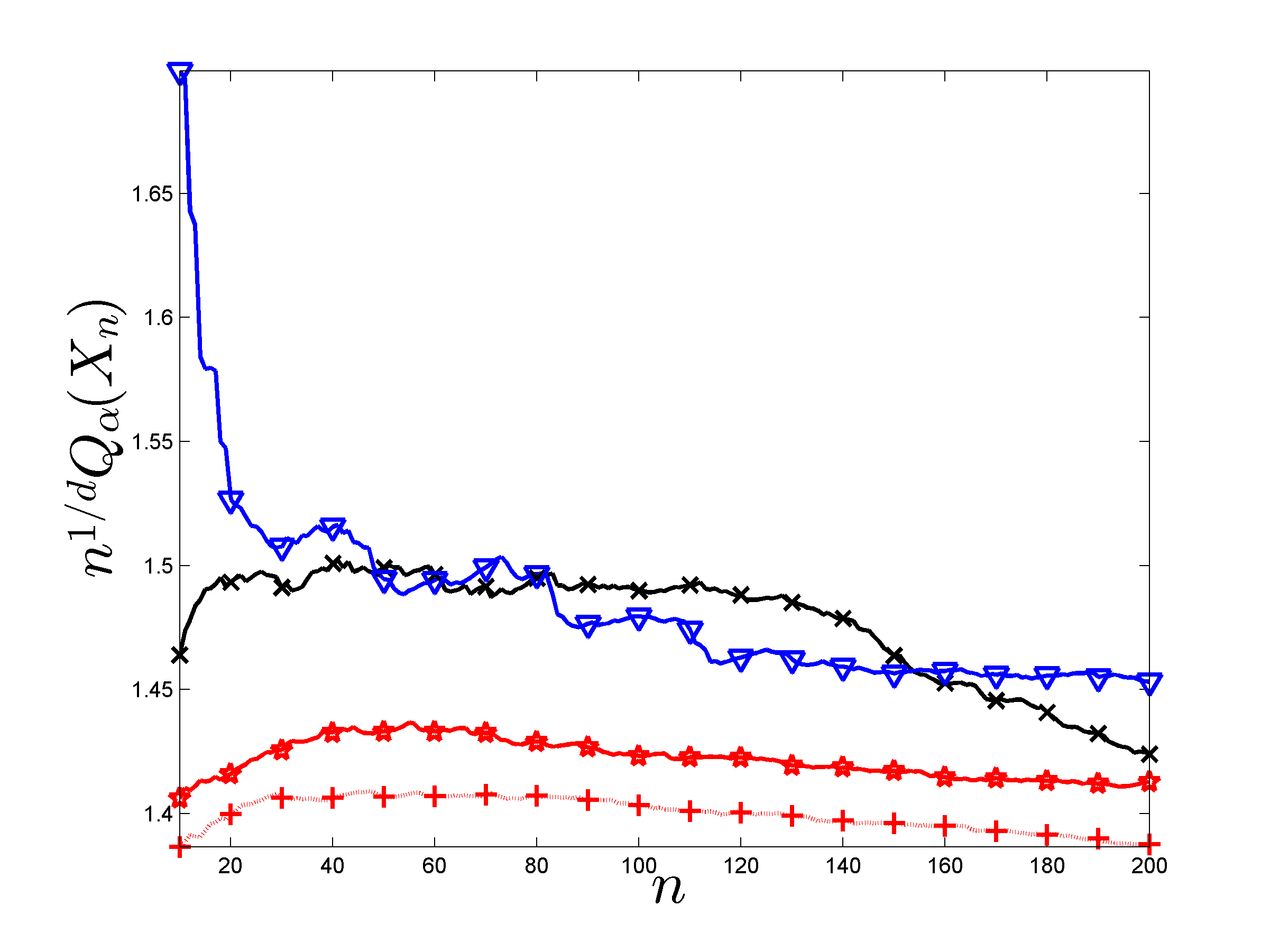}
\end{center}
\caption{Left: $\CR(\Xb_n)/R_\star(n_{\max},d)$; Right: $n^{1/d}\,Q_\ma(\Xb_n)$, $n=10,\ldots,200$, $d=10$: $\Xb_n^\star$ (red $\bigstar$, $q=10$), $\Hb_{n,10}$ (blue $\triangledown$), $\Sb_{n,10}$ (black~$\times$). The dotted red lines with $+$ correspond to $\Xb_n^\star$ when $\SX_Q$ is restricted to $\Sb_{Q',d}$.}
\label{F:LDS-d10}
\end{figure}

(\textit{i}) Figure ~\ref{F:LDS-d10} shows the comparison of $\Xb_n^\star$ (red) with $\Hb_{n,d}$ (blue $\triangledown$) and $\Sb_{n,d}$ (black $\times$). To illustrate the impact of the presence of the vertices of $\SC_d$ in $\SX_Q$, the performance of the c.d.f.-based method is presented both for $\SX_Q=\Sb_{Q',d}\cup\{0,1\}^d$ (solid red line with $\bigstar$) and $\SX_Q=\Sb_{Q',d}$ (dotted red lines with $+$).
The left panel plots the evolution of $\CR(\Xb_n)$, showing that the greedy maximization of $\widehat I_{B^\star,q}(\Zb_n)$ yields designs with significantly smaller covering radii than the two classical LDS. The right panel shows the evolution of $Q_\ma(\Xb_n)$, revealing an even stronger dominance of $\Xb_n^\star$ for this criterion. Notice the significant increase of $\CR$ when the vertices of $\SC_d$ are not included in $\SX_Q$ (dotted red line). Contrariwise, this results in a smaller covering $\ma$-quantile, since the empirical c.d.f.\ criterion ignores then the immediate neighborhoods of the vertices.

\begin{figure}[ht!]
\begin{center}
 \includegraphics[width=.49\linewidth]{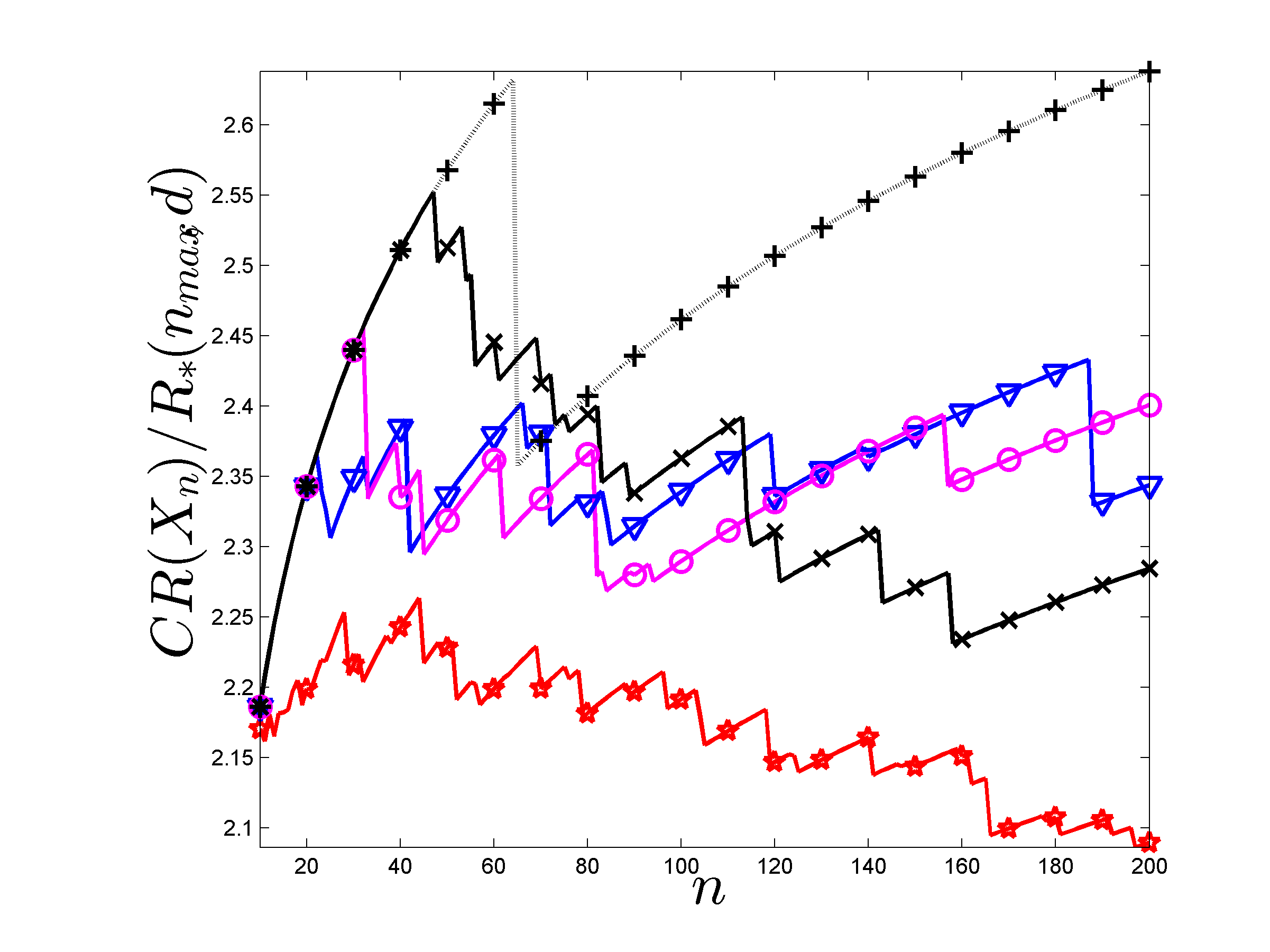} \includegraphics[width=.49\linewidth]{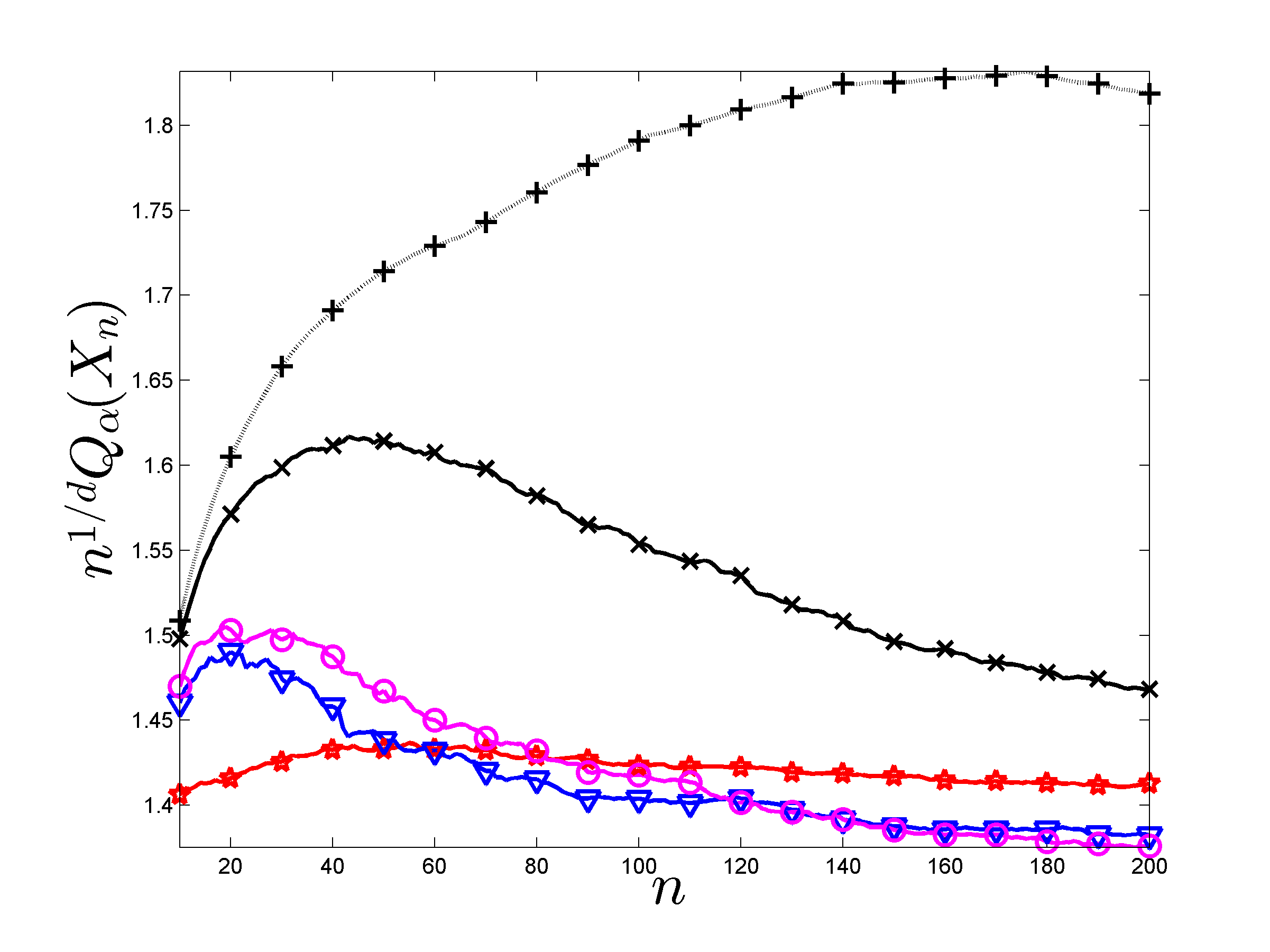}
\end{center}
\caption{Left: $\CR(\Xb_n)/R_\star(n_{\max},d)$; Right: $n^{1/d}\,Q_\ma(\Xb_n)$, $n=10,\ldots,200$, $d=10$: $\Xb_n^\star$ (red $\bigstar$, $q=10$), $\Xb_n^{CH,2\,\sqrt{2\,d}}$ (magenta $\circ$), $\Xb_n^{CH,\beta_\star(200,d)}$ with $\beta_\star(n,d)$ given by \eqref{beta*} (blue $\triangledown$) and $\Xb_n^{CH,\infty}$ (black $\times$, and dotted line with black $+$ for the candidate set $\Sb_{4\,096,10}\cup\{0,1\}^d$).}
\label{F:CH-d10}
\end{figure}

(\textit{ii}) Figure~\ref{F:CH-d10} shows the comparison of $\Xb_n^\star$ (red $\bigstar$, as in (\textit{i})) with the designs $\Xb_n^{CH,\beta}$ produced by three variants of coffee-house. The following code is used: $\beta=\infty$ (standard coffee-house design) in black $\times$, $\beta=2\,\sqrt{2\,d}$ in magenta $\circ$, and $\beta = \beta_\star(100,d)$ given by \eqref{beta*} in blue $\triangledown$. To show the sensitivity of standard coffee-house with respect to the candidate set, we also plot
in this figure (black +) the performance of coffee-house design $\Xb_n^{CH,\infty}$ for a candidate set extended by the vertices of $\SC_{10}$, $\SX_C=\Sb_{4\,096,10}\cup\{0,1\}^d$. The design $\Xb_n^\star$ has smaller $\CR$ (see left panel) than all versions of coffee-house for all design sizes considered\footnote{Our c.d.f.-based method thus inherits the ranking established in \cite{ShangA2020} where it is shown that $\Xb_n^{CH,2\,\sqrt{2\,d}}$ performs better in terms of $\CR$ than sliced and nested Latin hypercube designs proposed in the literature, which are only batch-incremental (see the references in \cite{ShangA2020}).}. Except for very small design sizes, the boundary-phobic versions of coffee-house have better $Q_\ma$ values (right panel) than $\Xb_n^\star$. Note that $\Xb_n^{CH,\infty}$ has much larger $Q_\ma$ than its competitors and is very sensitive to the choice of the candidate set: when $\SX_C$ includes the vertices of $\SC_{10}$, those are systematically selected at all iterations considered ($n\leq 200$), resulting in both large $\CR$ and $Q_\ma$.

\begin{figure}[ht!]
\begin{center}
 \includegraphics[width=.49\linewidth]{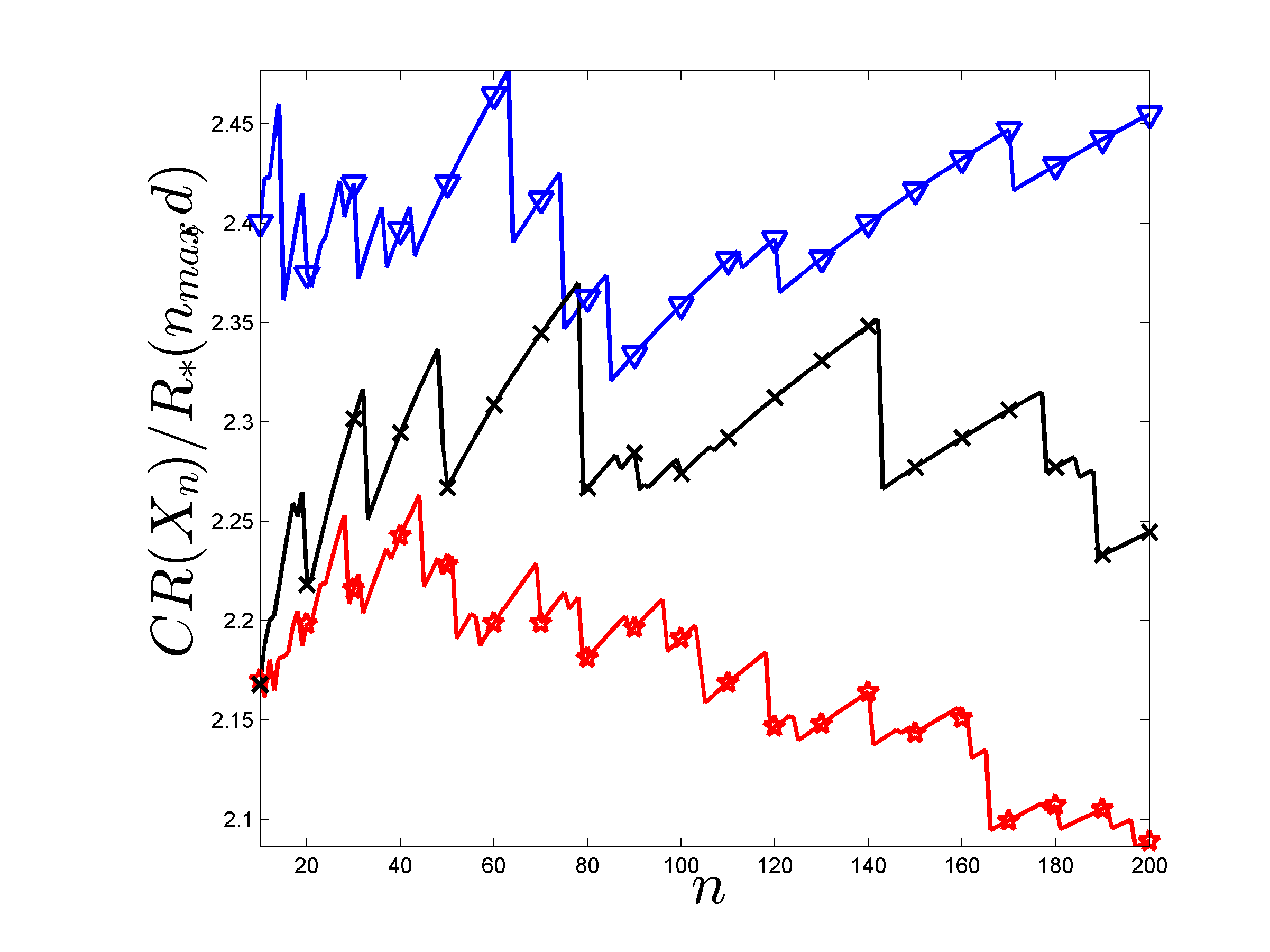} \includegraphics[width=.49\linewidth]{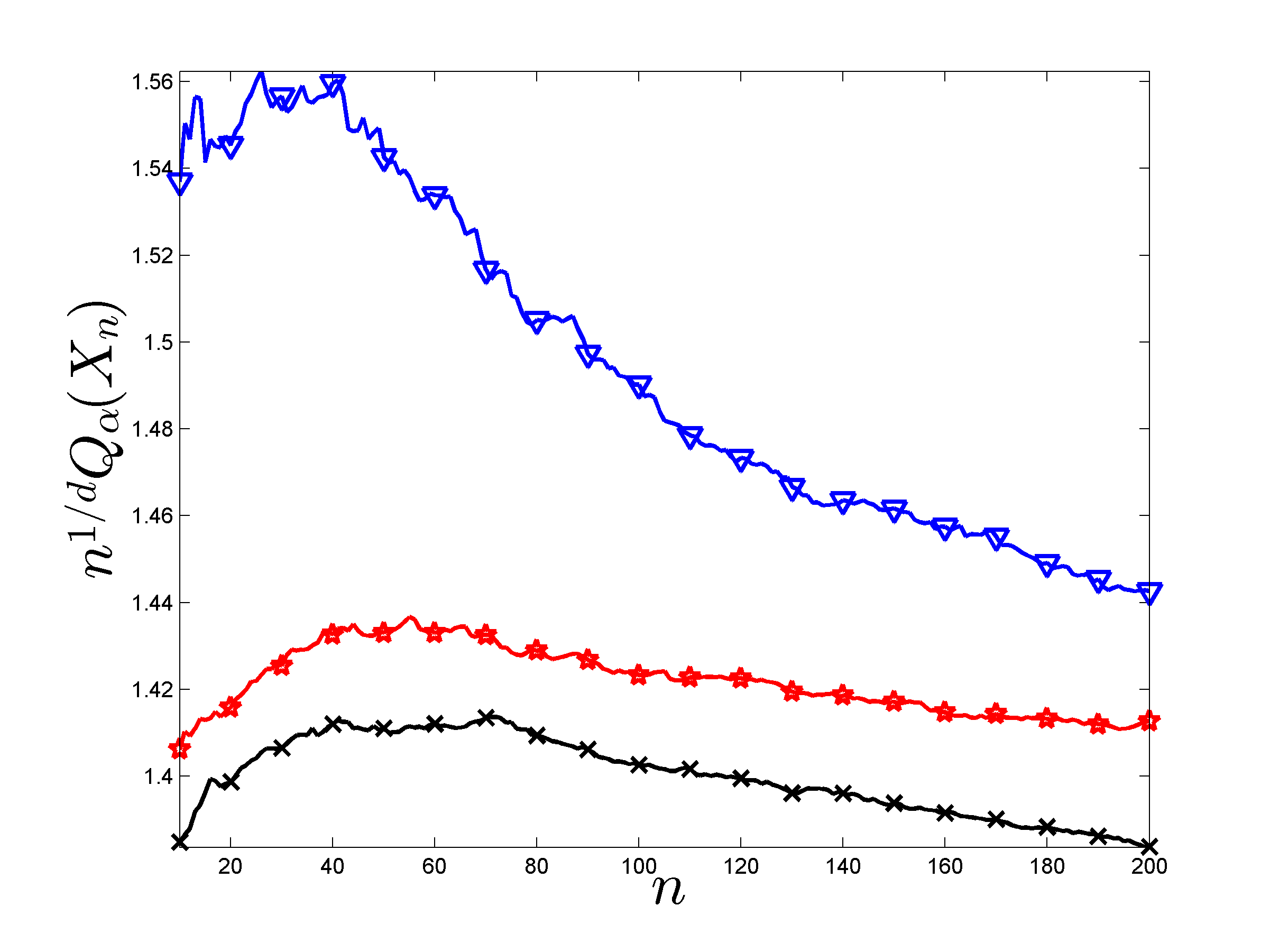}
\end{center}
\caption{Left: $\CR(\Xb_n)/R_\star(n_{\max},d)$; Right: $n^{1/d}\,Q_\ma(\Xb_n)$, $n=10,\ldots,200$, $d=10$; $\Xb_n^\star$ (red $\bigstar$), $\Xb_n^{VD}$ (blue $\triangledown$) and $\Xb_n^{RD}$ (black $\times$); $q=10$.}
\label{F:VD-RD-d10}
\end{figure}

(\textit{iii}) Figure~\ref{F:VD-RD-d10} shows the comparison of $\Xb_n^\star$ (red $\bigstar$, as in (\textit{i}) and (\textit{ii})) with the designs $\Xb_n^{VD}$ (blue $\triangledown$) and $\Xb_n^{RD}$ (black $\times$). As before we can see (left panel) that $\Xb_n^\star$ dominates $\Xb_n^{VD}$ and $\Xb_n^{RD}$ for $\CR$ for all design sizes. Note the very good performance of $\Xb_n^{RD}$ for the covering quantile (right panel).

\begin{figure}[ht!]
\begin{center}
 \includegraphics[width=.49\linewidth]{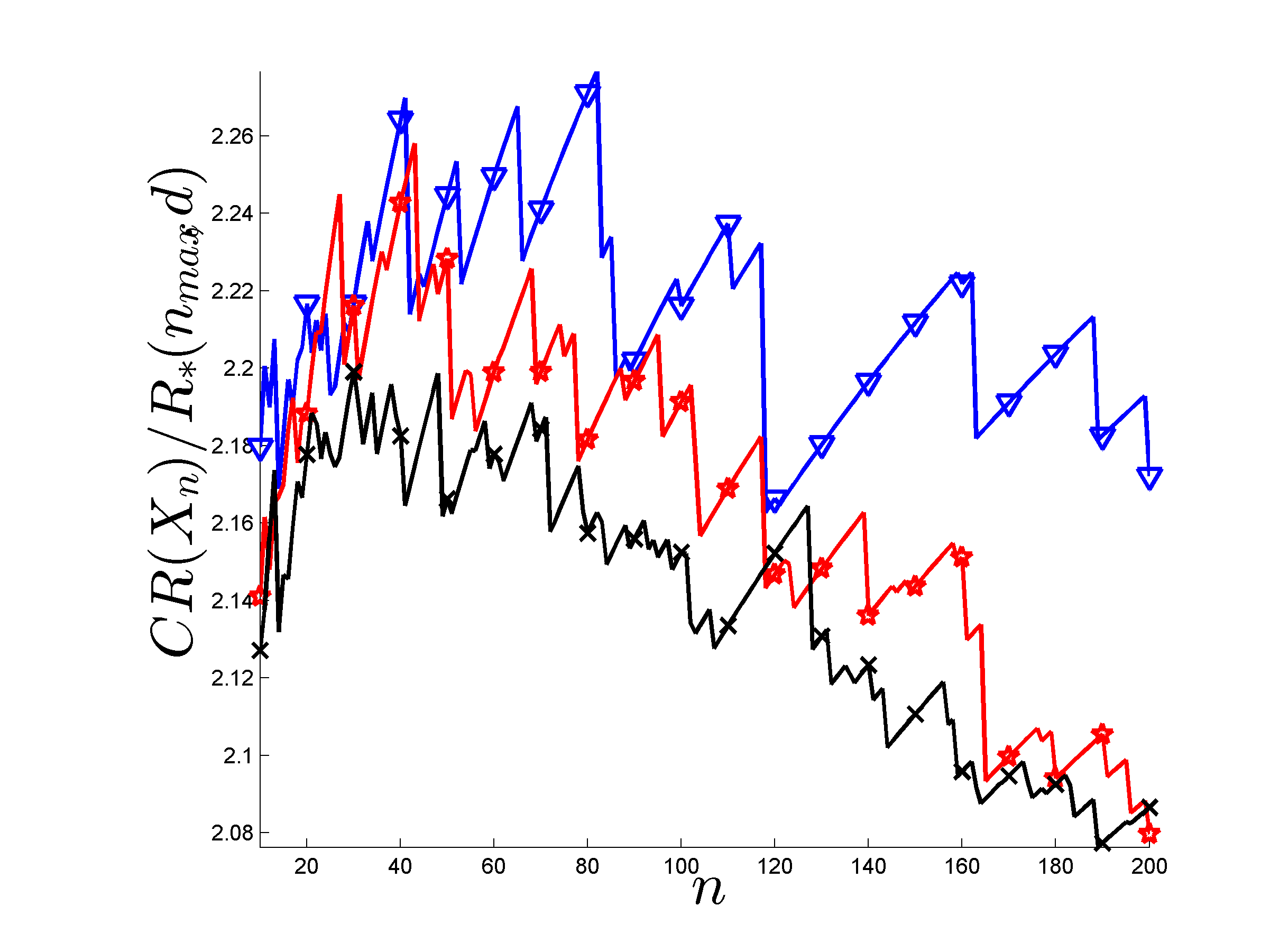} \includegraphics[width=.49\linewidth]{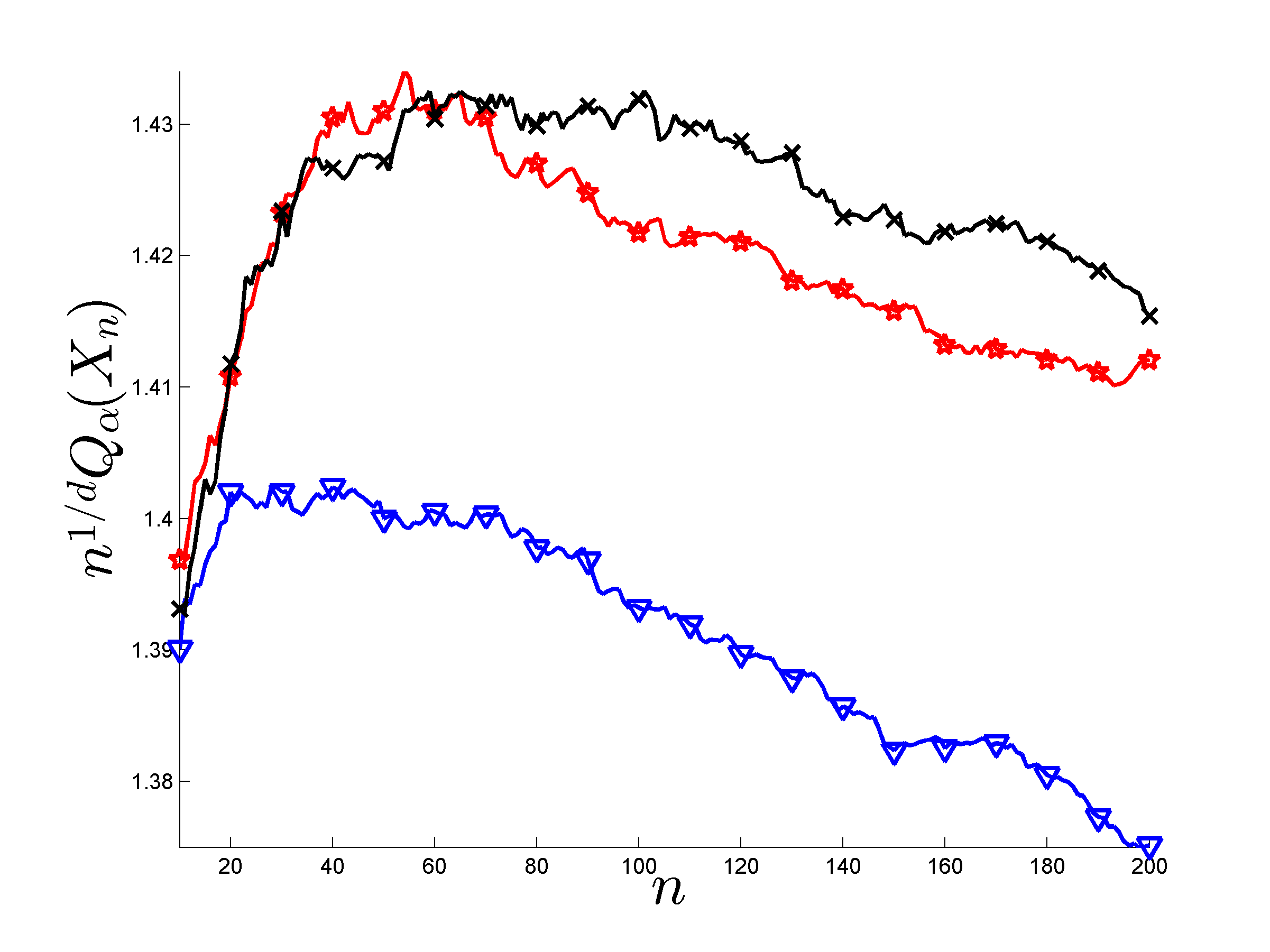}
\end{center}
\caption{Left: $\CR(\Xb_n^\star)/R_\star(n_{\max},d)$; Right: $n^{1/d}\,Q_\ma(\Xb_n^\star)$, $n=10,\ldots,200$, $d=10$;  $q=5$ (blue $\triangledown$), $q=10$ (red $\bigstar$), and $q=25$ (black $\times$).}
\label{F:q_5-10-25}
\end{figure}

Figure~\ref{F:q_5-10-25} confirms the robustness of the c.d.f.-based method with respect to $q$, displaying the evolution of both $\CR$ and $Q_\ma$ for $q=5$, $10$ and $25$. It shows that increasing $q$ reduces the covering radius of $\Xb_n^\star$ by giving more weight to points in $\SX_Q$ far from the design (i.e., the vertices of $\SC_d$), at the detriment of $Q_\ma$. Note the larger impact of increasing $q$ from 5 to 10 than from 10 to 25, suggesting that further increase of $q$ will not result in significantly different performance.

Table~\ref{Tb:CompTime-A-VR-RD} gives the computational time of methods considered, with $\Xb_n^{LRD}$ the design obtained by lazy-greedy minimization of $\Psi_q(\Zb_n,\mu_Q)$ given by \eqref{relaxed-CR}. The coffee-house variants are by far the fastest, followed by the c.d.f.-based method. Whenever numerical complexity must be kept to a minimum, one of the variants of coffee-house design based on spacings generates designs with slightly increased covering radii than $\Xb_n^\star$ but at a much smaller cost.

\begin{table}[ht!]
\caption{Computational time of $\Xb_{200}$ (in s, 10 repetitions).
}
\label{Tb:CompTime-A-VR-RD}
\begin{tabular}{ccccccc}
\hline\noalign{\smallskip}
$\Xb_n^\star$ & $\Xb_n^{VD}$ & $\Xb_n^{RD}$ & $\Xb_n^{LRD}$ & $\Xb_n^{CH,\infty}$ & $\Xb_n^{CH,2\,\sqrt{2\,d}}$ & $\Xb_n^{CH,\beta_\star(100,d)}$ \\
\noalign{\smallskip}\hline\noalign{\smallskip}
18.0 & 32.9 &  303.8 &  126.2 & 0.5 &  0.9 & 0.9 \\
\noalign{\smallskip}\hline
\end{tabular}
\end{table}

We assess now the impact of properties of the set of candidate points by repeating the comparison between $\Xb_n^\star$ and the coffee-house designs $\Xb_n^{CH,\beta}$ when $\SX_C=\Zb_{Lh,100}$, a Latin hypercube design with maximum packing radius; see \href{https://spacefillingdesigns.nl/}{https://spacefillingdesigns.nl/}. We define in this way an order in $\Zb_{Lh,100}$, hoping that all prefix designs $\Zb_{Lh,1:n}=\{\zb_i\}_{i=1}^n$, $\zb_i\in\Zb_{Lh,100}$, $n\leq 100$, have small covering radius.

\begin{figure}[ht!]
\begin{center}
 \includegraphics[width=.49\linewidth]{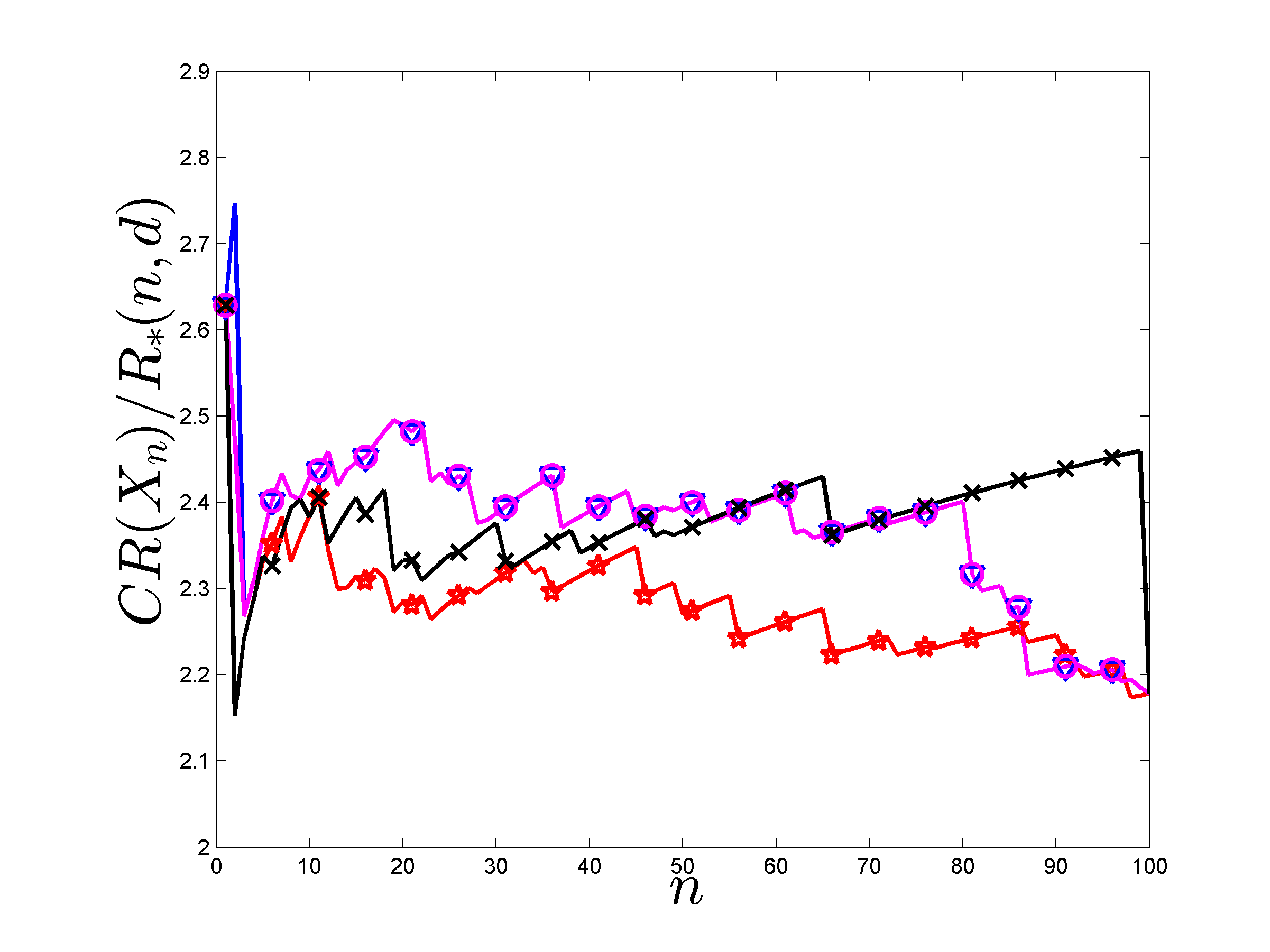} \includegraphics[width=.49\linewidth]{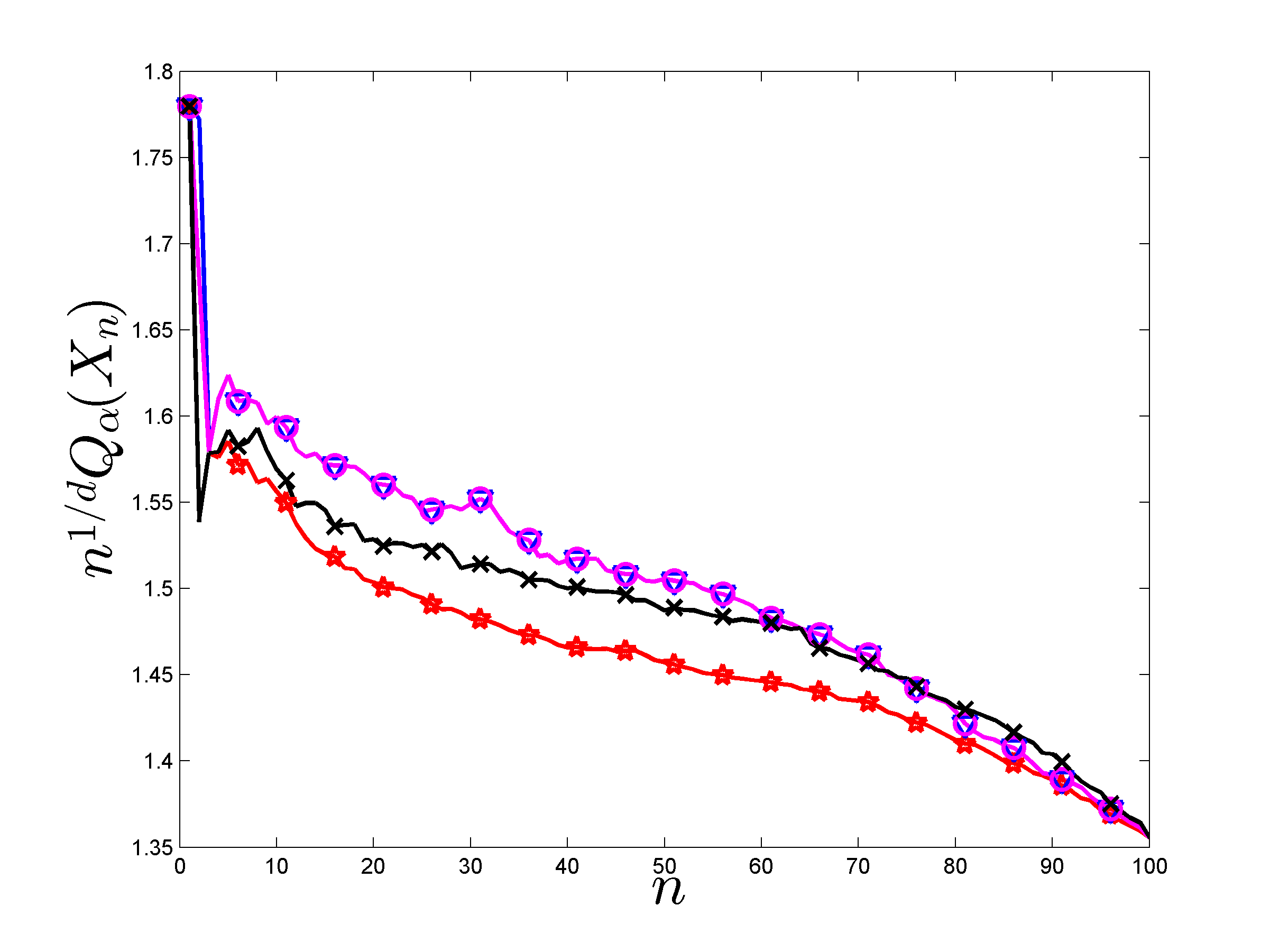}
\end{center}
\caption{Left: $\CR(\Xb_n)/R_\star(n_{\max},d)$; Right: $n^{1/d}\,Q_\ma(\Xb_n)$, $n=1,\ldots,n_{\max}=100$, $d=10$, candidate set $\SX_{100}=\Zb_{Lh,100}$: $\Xb_n^\star$ (red $\bigstar$, $q=10$) and $\Xb_n^{CH,\beta}$ with $\beta=2\,\sqrt{2\,d}$ (magenta $\circ$), $\beta=\beta_\star(200,d)$ given by \eqref{beta*} (blue $\triangledown$) and $\beta=\infty$ (black $\times$).}
\label{F:CR_d10_Lh}
\end{figure}

Figure~\ref{F:CR_d10_Lh} shows $\CR$ (left panel) and $Q_{\ma}$ (right panel) of designs $\Xb_n^\star$ along with those of $\Xb_n^{CH,\beta}$. The figure shows that for this extremely constrained candidate set $\Xb_n^\star$ (red) still outperforms the other methods\footnote{The magenta and blue curves, corresponding to $\beta=2\,\sqrt{2\,d}$ and $\beta = \beta_\star(100,d)$, coincide.}. This example confirms that the incremental optimization of the c.d.f.-based criterion produces, when applied to a final design with good space-filling properties, a sequence of nested designs with good overall space-filling properties (notice however that better $\CR$ values are obtained without the Latin hypercube constraint).

\bigskip
\paragraph{Example~2.}

We investigate in this example the impact of non-convex domains $\SX$, by considering an annular geometry, letting $\SX=\{\xb\in\mathds{R}^2: 1/2 \leq \|\xb\|\leq 1\}$ (for simplicity, we still base the covering and packing radii on the Euclidean distance, although geodesics could have been used as well). We compare the maximization of the c.d.f.-based criterion to (\textit{i}) use of prefixes of LDS and (\textit{ii}) to coffee-house design with $\beta=\infty$.
We take $\SX_C=\SX_Q$ given by the first $C=2\,048$ points of a Sobol' sequence of points in $\SC_2$  --- renormalized to $[-1,1]^2$ --- falling inside $\SX$. An analogous construction is used for the set $\SX_N$ used to approximate $\CR(\Zb_n)$, by retaining the first $N=2^{18}$ points of a renormalized scrambled Sobol' sequence that fall inside $\SX$. The LDS design $\Sb_{n,d}$ gathers the first $n$ elements of $\SX_C$.

Spacings are  difficult to handle for non-convex domains, explaining why only $\beta=\infty$ is studied in this example. Instead, we consider two distinct  candidate sets in the construction of coffee-house designs: $\SX_C$, the points of the Sobol' renormalized and clipped  sequence, and, in an effort to enforce  boundary avoidance, we also use a different candidate set  $\SX'_C$ which is the restriction of the renormalized Sobol' sequence to the eroded annulus $\SX'=\{\xb\in\mathds{R}^2: 1/2+r\leq \|\xb\|\leq 1-r\}\subset\SX$, with $r=R_\star(100,d)/2$.

\begin{figure}[ht!]
\begin{center}
 \includegraphics[width=.49\linewidth]{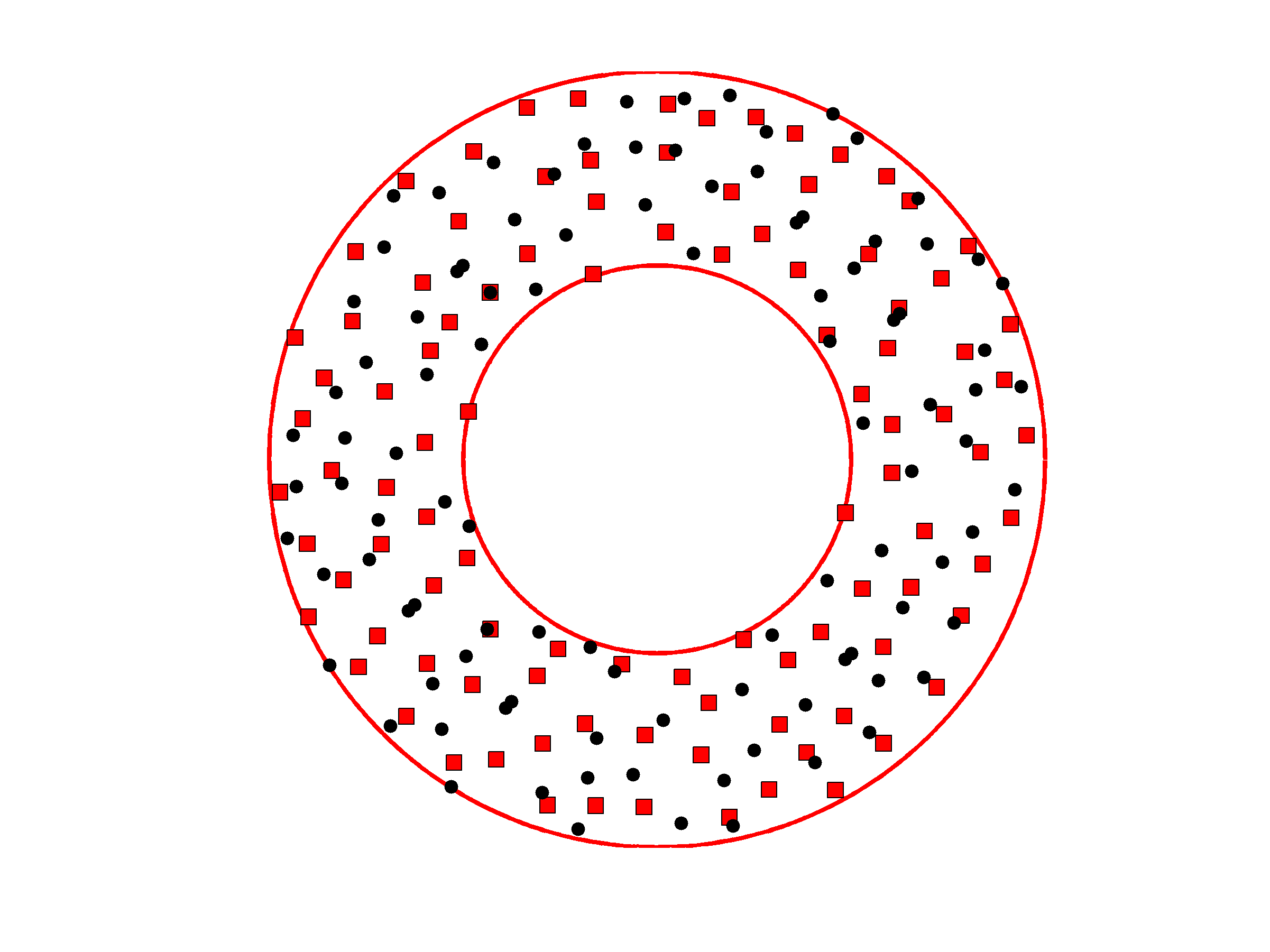} \includegraphics[width=.49\linewidth]{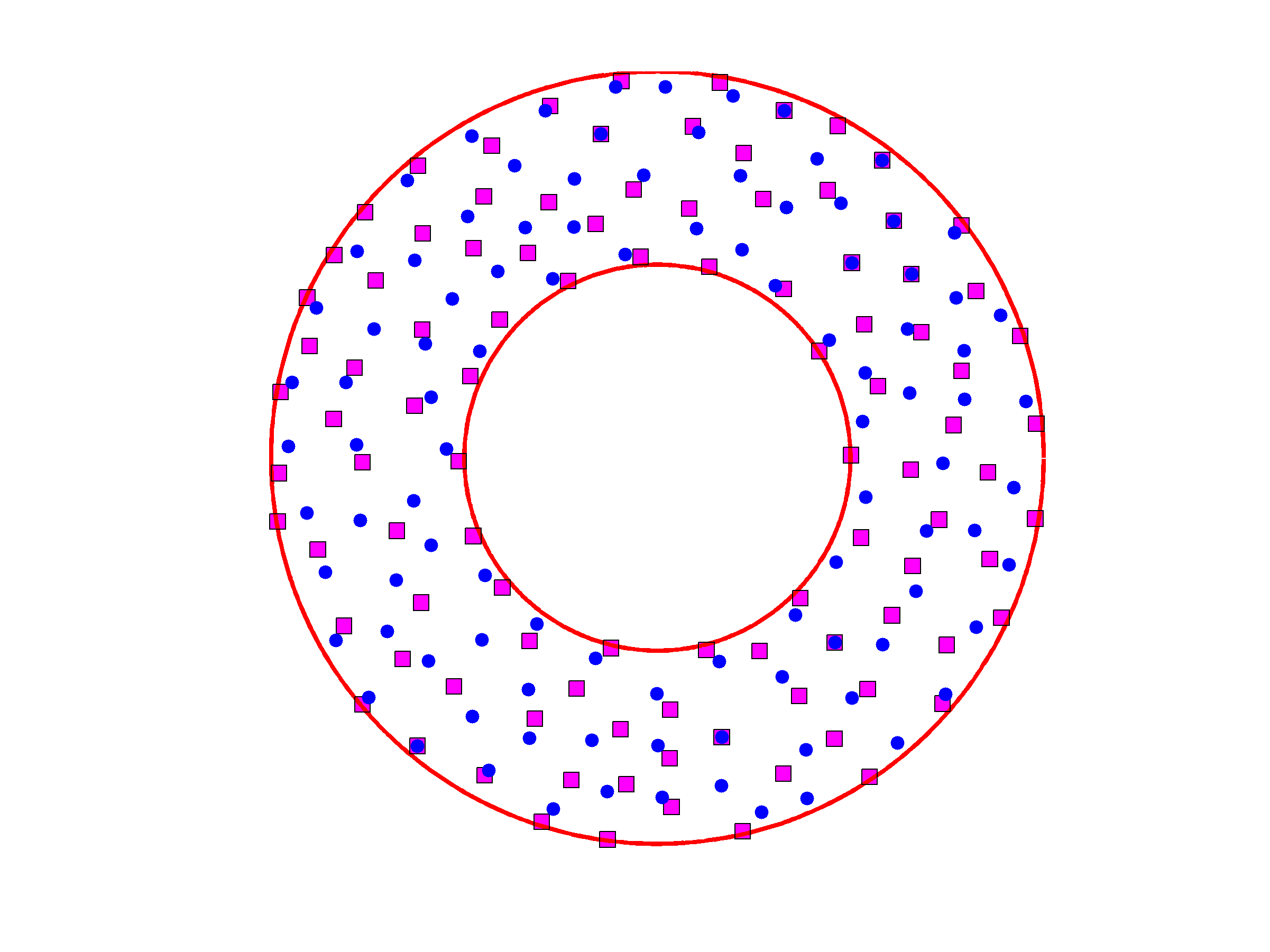}
\end{center}
\caption{Left: $\Xb_n^\star$ (red $\blacksquare$) and $\Sb_{n,2}$ (black $\bullet$); Right: $\Xb_n^{CH,\infty}$ (magenta $\blacksquare$) and $\Xb_n^{CH',\infty}$ (blue $\bullet$).}
\label{F:X_50_100_annulus}
\end{figure}

Figure~\ref{F:X_50_100_annulus} shows the 100-point designs obtained. The left panel shows $X^\star_n$ (red $\blacksquare$) and $\Xb_n^S$ (black $\bullet$), and the right panel displays the coffee-house designs $\Xb_n^{CH,\infty}$  (magenta $\blacksquare$) and  $\Xb_n^{CH',\infty}$  (blue $\bullet$) obtained for the candidate sets $\SX$ and  $\SX'$, respectively. Visual inspection shows that they all fill $\SX$ reasonably well, although $\Sb_{n,d}$ has a few pairs of nearly coincident points.

\begin{figure}[ht!]
\begin{center}
 \includegraphics[width=.49\linewidth]{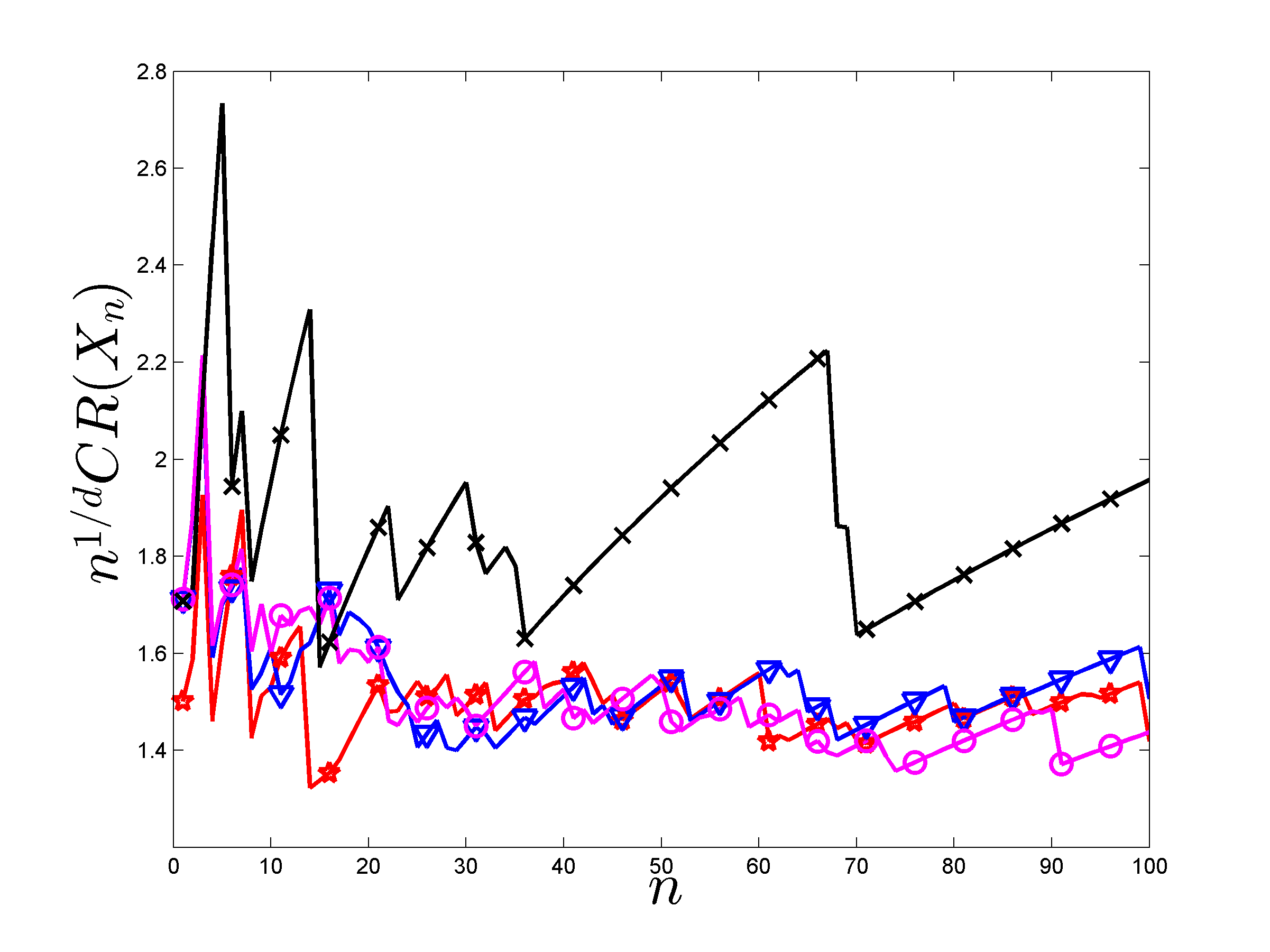} \includegraphics[width=.49\linewidth]{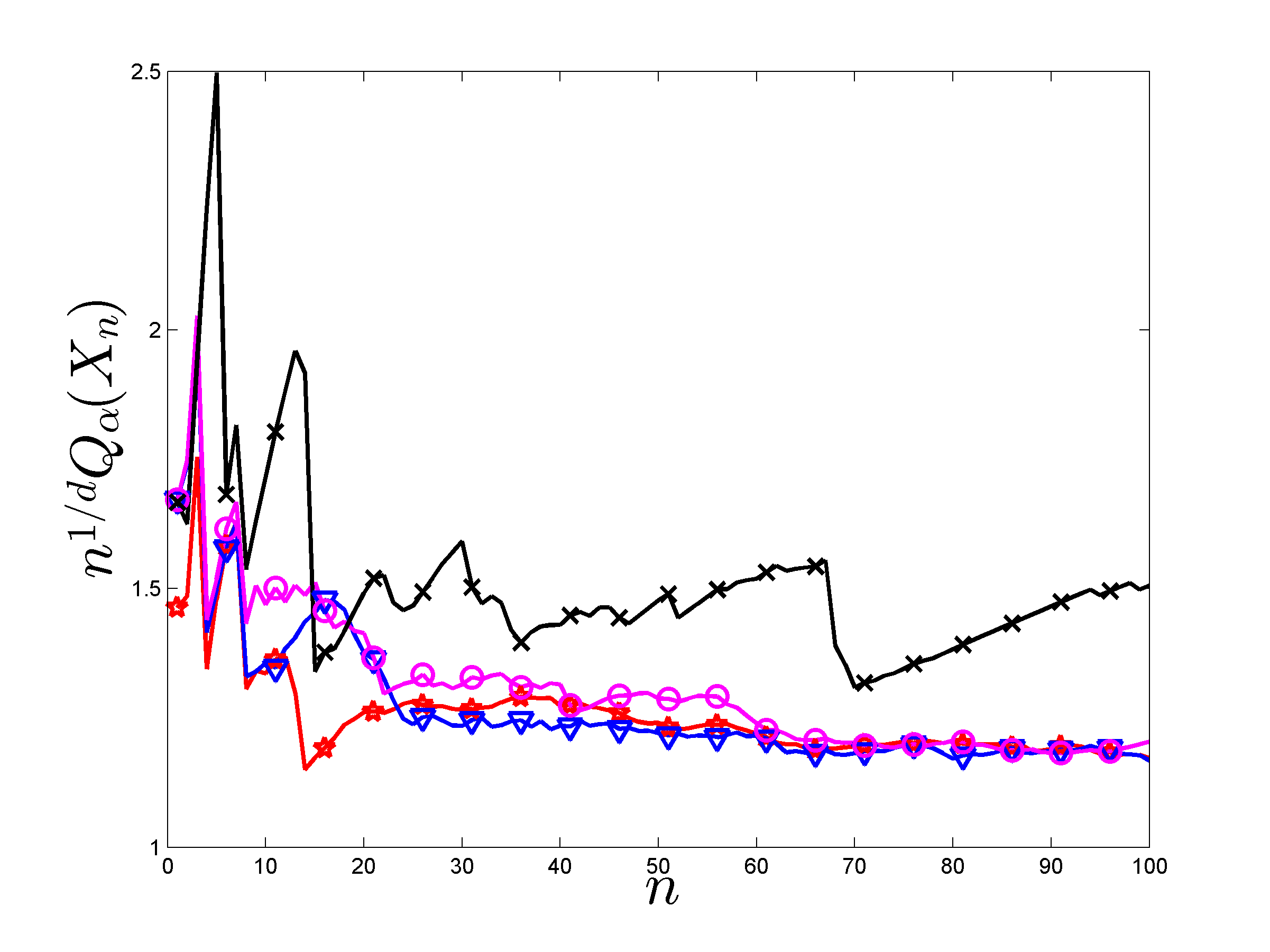}
\end{center}
\caption{$n^{1/d}\,\CR(\Xb_n)$ (left) and $n^{1/d}\,Q_{\ma,[\SX_N]}(\Xb_n)$ (right) for the four designs in Figure~\ref{F:X_50_100_annulus}: $\Xb_n^\star$ (red $\bigstar$), $\Sb_{n,2}$ (black $\times$), $\Xb_n^{CH,\infty}$ (magenta $\circ$) and $\Xb_n^{CH',\infty}$ (blue $\triangledown$).}
\label{F:CR-rho_50_100_annulus}
\end{figure}

Figure~\ref{F:CR-rho_50_100_annulus} shows the performance of the four designs, in terms of normalized $\CR$ (left panel) and $Q_\ma$ (right panel). The three greedy constructions exhibit comparable performance and clearly outperform the Sobol' sequence for both criteria. $\Xb_n^\star$ and $\Xb_n^{CH,\infty}$ have slightly smaller covering radii than $\Xb_n^{CH',\infty}$ for $n \gtrsim 50$: the absence of sharp corners and edges seems to soften the importance of staying away from the boundaries of the domain; notice that $\Xb_n^{CH,\infty}$ has many points near the boundary while $\Xb_n^\star$ has few, suggesting that boundaries are not necessarily the main concern here.

\section{Conclusions}

This paper proposes a novel space-filling c.d.f.-based criterion $\IBq$ tailored to the minimization of the covering radius. This set function satisfies the conditions of the celebrated Nemhauser's Theorem on the greedy maximization of submodular functions, thereby opening the possibility to challenge the simple but 50\%-efficient coffee-house design methodology.  Although the performance guarantees provided by the submodularity of $\IBq$ do not translate into performance guarantees for the covering radius, we verified experimentally, over a larger set of case studies than those presented in the paper, that the design sequences obtained by its greedy maximization have smaller $\CR$ values than
competing state-of-the-art incremental design methods. The fact that it almost always outperforms coffee-house which has a guaranteed 50\% efficiency suggests that an even stronger guarantee may hold for our method.

By fully taking into account how the points of $\Xb_n$ are collectively embedded in the domain $\SX$, $\IBq$ overcomes the limitations due to the locality of $\CR(\Xb_n)$ mentioned in the introduction. It involves a regularization parameter $q$ and a range parameter $B$, whose precise values do not significantly impact its ability to express the space-fillingness of a design. The method also requires the specification of a set of points $\SX_Q$ to compute the distance c.d.f.\ and of a candidate set $\SX_C$ of eligible design points. For finite $q$, the method greedily minimizes the $L^{q+1}$-mean quantization error evaluated on $\SX_Q$, which, when $B=\diam(\SX)$ and $q\to\infty$, tends to an underestimate of $\CR(\Xb_n)$ (since based on $\SX_Q$ only). We verified that setting $q\in[5,25]$ and $B=\diam(\SX)$ yields good designs in a wide variety of situations. The choice of $\SX_C$ is not critical, even in the challenging situation where $\SX$ is the hypercube $\SC_d$, for which the presence of vertices is easily problematic for large $d$: the method never tries to select points near the vertices and, more generally, is not sensitive to the proximity of points of $\SX_C$ to the boundary of $\SX$.

The coffee-house algorithm only requires the specification of a candidate set $\SX_C$ and runs much faster. However, when $\SX=\SC_d$ it naturally tends to select points as near as possible to the vertices of $\SX$, thereby exhibiting a lack of robustness to the choice of $\SX_C$. Boundary avoidance can be enforced through the notion of spacings, resulting in simple and efficient algorithms when $\SX$ has a simple enough geometry.

Overall, the geometry of $\SX$ appears as being a key factor, including in terms of the relevance of $\CR$ to measure the space-filling quality of a design. We thus concur with the authors of \cite{NoonanZ2020} on the importance of using a more global characteristic like the covering $\ma$-quantile $Q_\ma$. Incremental design constructed with $\IBq$ are also competitive in terms of $Q_\ma$ compared to other incremental design methods.

Note finally that the development of greedy algorithms for the minimization of $Q_\ma(\Xb_n)$ itself forms another challenging but promising objective towards the incremental construction of efficient space-filling designs, in particular in the hypercube $\SC_d$.

\vspace{0.5cm}
\appendix
\vsp
\noindent{\Large\bf Appendix}

\paragraph{Two-point $\CR$-optimal design in $\SC_d$.}
The $\CR$-optimal one-point design is $\Zb_1^\star=\1b_d/2$, i.e., the center of $\SC_d$.

Consider the design $\Zb_2(\ma)=\{\zb_1(\ma),\zb_2(\ma)=\1b_d-\zb_1(\ma)\}$ with $\zb_1(\ma)=(1/2,\ldots,1/2,\ma)$, $\ma< 1/2$. It defines a partition of $\SC_d$ into two polyhedral Voronoi cells $\SC_1$ and $\SC_2$ which are separated by the bisecting hyperplane $\SH$ of the segment joining $\zb_1(\ma)$ and $\zb_2(\ma)$, $\SH=\{\zb\in \SC_d: z_d=1/2\}$. We have $\CR(\Zb_2(\ma))=\|\xb^\star-\zb_1(\ma)\|$ for $\xb^\star$ a vertex of $\SC_d$ when $\ma\geq 1/4$ and for $\xb^\star$ having all its coordinates in $\{0,1\}$ except the last one equal to $1/2$ otherwise, implying $\CR(\Zb_2(\ma)) = \max\{[(d-1)/4+\ma^2]^{1/2},[(d-1)/4+(1/2-\ma)^2]^{1/2}\}$, whose minimum value is reached for $\ma=1/4$, with $\CR(\Zb_2(1/4))=(1/2)\,\sqrt{d-3/4} < \sqrt{d}/2$.

Let $\Zb_2^\star=\{\zb_1^\star,\zb_2^\star\}$ denote a 2-point $\CR$-optimal design, and denote by $\SC_1^\star$ and $\SC_2^\star$ the two cells of the Voronoi tessellation of $\SC_d$ that $\Zb_2^\star$ generates. Each of them contains exactly $2^{d-1}$ vertices of $\SC_d$, since otherwise one of them, $\SC_1^\star$ say, would contain two opposite vertices, implying $\CR(\Zb_2^\star) \geq \diam(\SC_1^\star)/2 = \sqrt{d}/2 > \CR(\Zb_2(1/4))$. We have $\CR(\Zb_2^\star) \geq \max\{\diam(\SC_1^\star)/2,\diam(\SC_2^\star)/2\}$. The minimum is obtained when the two cells have the same diameter and are such that this diameter is minimal, which gives $\CR(\Zb_2^\star)=\CR(\Zb_2(1/4))$.

Consider now an incremental greedy construction, one-step-ahead optimal. Since any 2-point design of the form $\Zb_2=\{\1b_d/2,\zb\}$ is such that $\CR(\Zb_2)=\sqrt{d}/2$, we necessarily have $\CR(\Xb_2)=\sqrt{d}/2>\CR(\Zb_2^\star)$. Note that if we could add two points at a time, we could reach $\CR(\{\1b_d/2,\xb_2,\xb_3\})=\CR(\Zb_2^\star)$ by taking $\xb_2=\zb_1^\star$, $\xb_3=\zb_2^\star$, and even $\CR(\{\1b_d/2,\widehat\xb_2,\widehat\xb_3\})=(1/2)\,\sqrt{d-8/9}$ by choosing $\widehat\xb_2=(1/2,\ldots,1/2,1/6)$ and $\widehat\xb_3=\1b_d-\widehat\xb_2$.

\paragraph{Evaluation of $\CR(\xb_n)$ when $\SX=\SC_d$.}

\begin{figure}[ht!]
\begin{center}
\includegraphics[width=.49\linewidth]{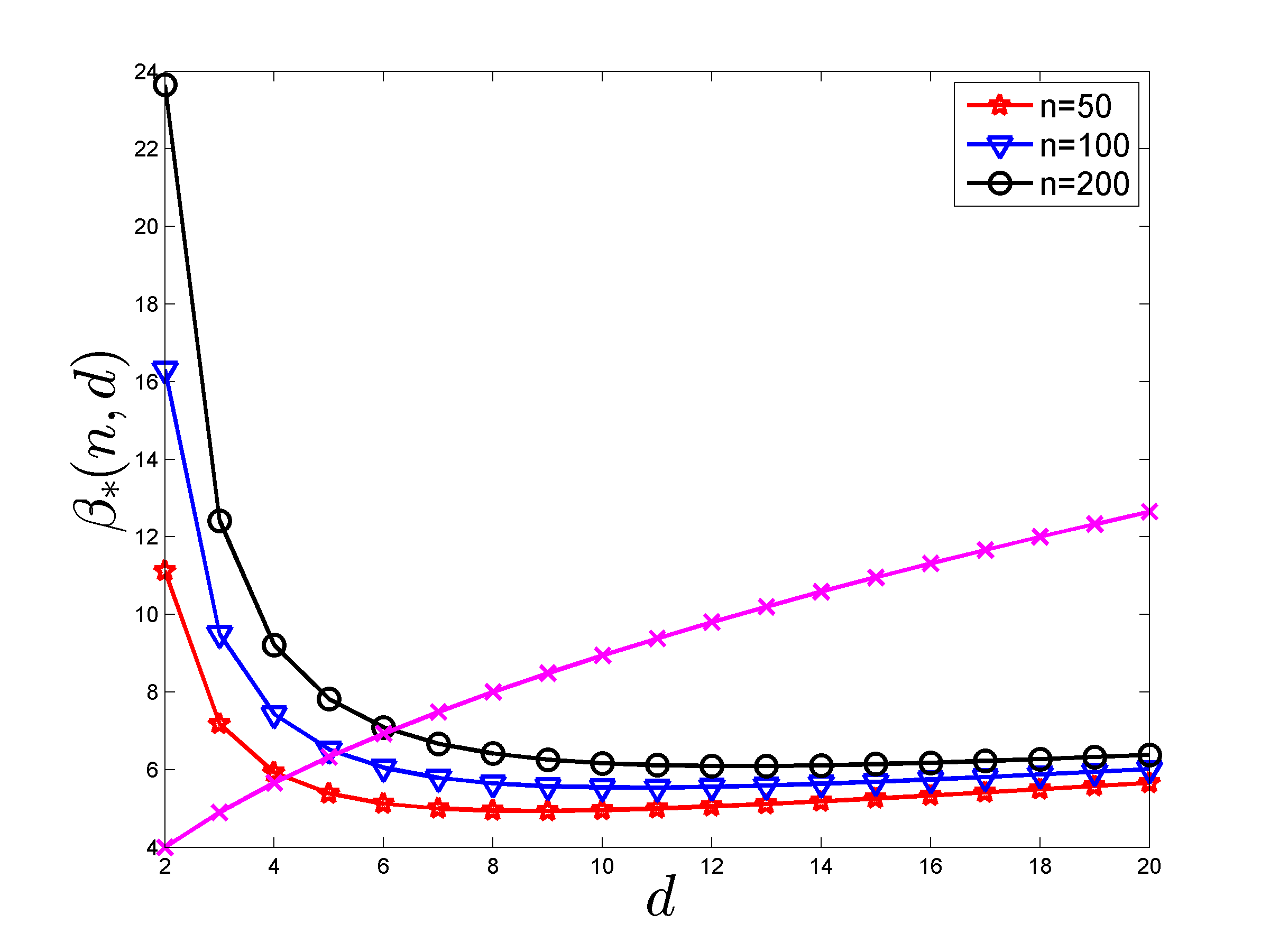}
\includegraphics[width=.49\linewidth]{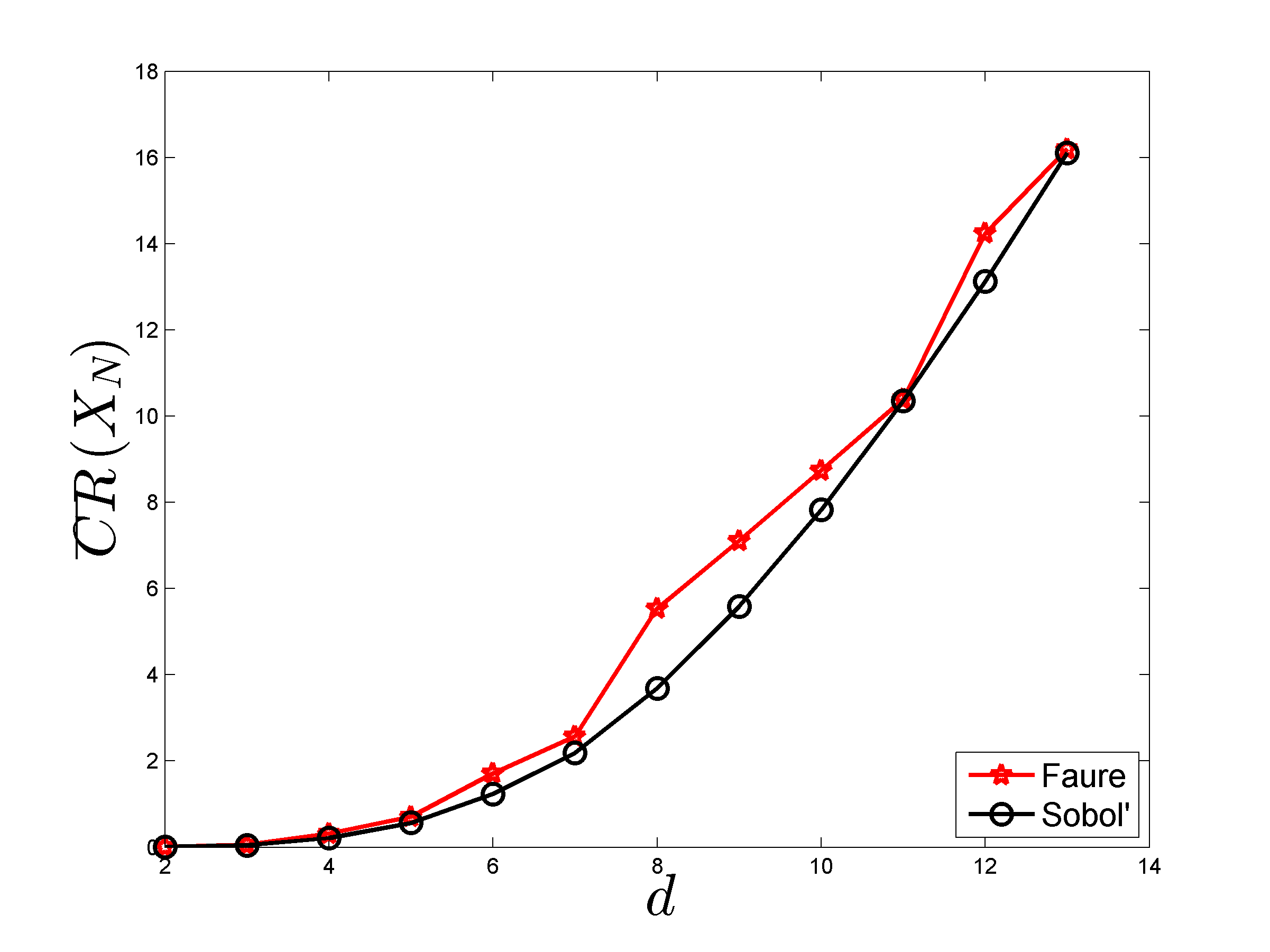}
\end{center}
\caption{Left: $\beta_\star(n,d)$ given by \eqref{beta*} as a function of $d$ for
$n=50, 100, 200$; the magenta curve with $\times$ is for the value $\beta=2\,\sqrt{2\,d}$ suggested in \cite{ShangA2020}.
Right: upper bound $\overline{\CR}(\SX_N)$ given by \eqref{upper-bound-CRN} as a function of $d$ for the sequences of Faure and Sobol' for $N=10^6$.}\label{F:CR_bound_Faure_Sobol}
\end{figure}

For $d\leq 4$, the exact values of $\CR(\Xb_n)$ are calculated using Voronoi tessellations; see \cite{P-JSFdS2017}; for larger $d$, we underestimate $\CR(\Xb_n)$ by $\CR_{[\SX_N]}(\Xb_n)$ given by \eqref{CR-approx}, with $\SX_N$ a finite subset of $\SX$. When $\SX_N$ corresponds to the first $N$ points of a $(t,s)$-sequence in base $\beta$ in $\SX$, we have
\be\label{upper-bound-CRN}
\CR(\SX_N) < \overline{\CR}(\SX_N) = \frac{\sqrt{d}\, \beta^{1+t/d}}{N^{1/d}} \,,
\ee
see \cite[Th.~6.11]{Niederreiter92}. The values of $t$ for the sequences of Sobol' (for which $\beta=2)$ are given in Table~\ref{Tb:Sobol}; see \cite{Sobol1967}. The sequences of Faure \cite{Faure1982} have $t=0$ but $\beta$ is the first prime number larger than or equal to $d$, making the upper bound in \eqref{upper-bound-CRN} larger than for Sobol'. Figure~\ref{F:CR_bound_Faure_Sobol} shows $\overline{\CR}(\SX_N)$ as a function of $d$ for both types of sequences when $N=10^6$, showing that the inequality
$\CR(\Xb_n) \leq  \CR_{[\SX_Q]}(\Xb_n) + \sqrt{d}\, \beta^{1+t/d}\, N^{-1/d}$ implied by \eqref{ineq-CR-approx} is of limited practical use. For that reason, for $d>4$ we have chosen to only indicate the value of $\CR_{[\SX_N]}(\Xb_n)$ for $\SX_N$ given by the union of $\Sb_{N_S}$, the first $N_S$ points of a (possibly scrambled) Sobol' sequence, and a $2^d$ full factorial design (i.e., the vertices of $\SX$), with thus $N=N_S+2^d$.

\begin{table}[ht!]
\caption{Values of $t(d)$ for Sobol' sequence, for $2\leq d \leq 13$.}
\label{Tb:Sobol}
\begin{tabular}{lllllllllllll}
\hline\noalign{\smallskip}
d & 2 & 3 & 4 & 5 & 6 & 7 & 8 & 9 & 10 & 11 & 12 & 13 \\
\noalign{\smallskip}\hline\noalign{\smallskip}
t & 0 & 1 & 3 & 5 & 8 & 11 & 15 & 19 & 23 & 27 & 31 & 35 \\
\noalign{\smallskip}\hline
\end{tabular}
\end{table}

\paragraph{Proof of Theorem~\ref{TH:CH}.}
We first prove by induction that $P_\beta(\Xb_{n+1}) = (1/2)\,S_\beta(\Xb_n)$ for $n\geq 1$. Since $d(\xb_2,\partial\SX)\leq d(\xb_1,\partial\SX)$, we have
\bea
P_\beta(\Xb_2) = \frac12\, \min\left\{ \beta\,d(\xb_1,\partial\SX)\,, \ \beta\,d(\xb_2,\partial\SX)\,, \ \|\xb_1-\xb_2\| \right\} = \frac12\, \min\left\{\beta\, d(\xb_2,\partial\SX)\,, \ \|\xb_1-\xb_2\| \right\} = \frac12\, S_\beta(\Xb_1) \,,
\eea
which proves the property for $n=1$.
Assume that $P_\beta(\Xb_{n}) = (1/2)\,S_\beta(\Xb_{n-1})$. By construction, we have $P_\beta(\Xb_{n+1}) \leq (1/2)\,S_\beta(\Xb_n)$; we show that $P_\beta(\Xb_{n+1}) \geq (1/2)\,S_\beta(\Xb_n)$. The greedy construction of $\Xb_{n+1}$ implies that $d(\xb_{n+1},\Xb_n)\geq S_\beta(\Xb_n)$ and $\beta\,d(\xb_{n+1},\partial\SX)\geq S_\beta(\Xb_n)$. Also, any $\xb_i\neq\xb_j\in\Xb_n$ satisfy
\bea
\|\xb_i-\xb_j\| \geq 2\, P_\beta(\Xb_n) = S_\beta(\Xb_{n-1}) \geq S_\beta(\Xb_n) \,,
\eea
where the first inequality comes from the definition of $P_\beta$, the equality comes from the induction hypothesis and the second inequality from the fact that $S_\beta$ is nonincreasing. Similarly, $\beta\,d(\xb_i,\partial\SX) \geq 2\, P_\beta(\Xb_n) \geq S_\beta(\Xb_n)$ for all $\xb_i\in\Xb_n$. Altogether, this gives $P_\beta(\Xb_{n+1}) \geq (1/2)\,S_\beta(\Xb_n)$.

Let $\Zb_n^\star$ and $\Zb_n^{\star\star}$ be two $n$-point designs in $\SX$ such that $S_\beta(\Zb_n^\star)=S_{\beta,n}^\star=\min_{\Zb_n\subset\SX} S_\beta(\Zb_n)$ ($n\geq 1$) and $P_\beta(\Zb_n^{\star\star})=P_{\beta,n}^\star=\max_{\Zb_n\subset\SX} P_\beta(\Zb_n)$ ($n\geq 2$). Denote by $\SX^-(r)$ the erosion of $\SX$ by $\SB(\0b,r)$, $\SX^-(r)=\SX\ominus\SB(\0b,r)=\{\xb\in\SX: \xb+\zb\in\SX\,, \ \forall \zb\in\SB(\0b,r)\}$, with $\SX^-(r)\neq\emptyset$ for any $r\leq \diam(\SX)/2$.

Suppose that the $n$ balls $\SB(\xb_i,S_\beta(\Xb_n))$ do not cover $\SX^-(S_\beta(\Xb_n)/\beta)$. This would imply the existence of $\xb^\star\in\SX^-(S_\beta(\Xb_n)/\beta)$ such that $\beta\,d(\xb^\star,\partial\SX)>S_\beta(\Xb_n)$ and $d(\xb^\star,\Xb_n)>S_\beta(\Xb_n)$, and therefore $D_\beta(\xb^\star,\Xb_n)>S_\beta(\Xb_n)$. This is impossible since $S_\beta(\Xb_n)=\max_{\xb\in\SX} D_\beta(\xb^\star,\Xb_n)$.

We prove that $S_\beta(\Xb_n)\leq 2\, S_{\beta,n}^\star$.
Take any $\xb_i\in\Xb_{n+1}$. It satisfies $\xb_i\in\SX^-(2\, P_\beta(\Xb_{n+1})/\beta)$ from the definition of $P_\beta$, and thus  $\xb_i\in\SX^-(S_\beta(\Xb_n)/\beta)$ since $P_\beta(\Xb_{n+1}) = (1/2)\,S_\beta(\Xb_n)$, and $\xb_i\in\SX^-(S_{\beta,n}^\star/\beta)$ since $S_{\beta,n}^\star\leq S_\beta(\Xb_n)$.
Since the $n$ balls $\SB(\zb_i^\star,S_\beta(\Xb_n))$, $\zb_i^\star\in\Zb_n^\star$, cover $\SX^-(S_{\beta,n}^\star/\beta)$,
this implies the existence of a $\zb_\ell^\star\in\Zb_n^\star$ and of $\xb_j\neq\xb_k\in\Xb_{n+1}$ such that $\xb_j,\xb_k\in\SB(\zb_\ell^\star,S_{\beta,n}^\star)$. Therefore, $S_\beta(\Xb_n)=2\,P_\beta(\Xb_{n+1}) \leq \|\xb_i-\xb_j\| \leq 2\, S_{\beta,n}^\star$.

We now prove that $P_{\beta,n+1}^\star \leq 2\,P_\beta(\Xb_{n+1})$. Take any $\zb_i^{\star\star}\in\Zb_{n+1}^{\star\star}$; $\zb_i^{\star\star}\in\SX^-(2\, P_{\beta,n+1}^\star/\beta)$ from the definition of $P_\beta$, $\zb_i^{\star\star}\in\SX^-(2\, P_\beta(\Xb_{n+1})/\beta)=\SX^-(S_\beta(\Xb_n)/\beta)$ since $P_{\beta,n+1}^\star\geq P_\beta(\Xb_{n+1})=(1/2)\,S_\beta(\Xb_n)$, and $\zb_i^{\star\star}\in\SX^-(S_{\beta,n}^\star/\beta)$ since $S_{\beta,n}^\star\leq S_\beta(\Xb_n)$. Therefore, there exist $\zb_\ell^\star\in\Zb_n^\star$ and $\zb_j^{\star\star}\neq\zb_k^{\star\star}\in\Zb_{n+1}^{\star\star}$ such that $\zb_j^{\star\star},\zb_k^{\star\star}\in\SB(\zb_\ell^\star,S_{\beta,n}^\star)$. It implies that $2\, P_{\beta,n+1}^\star \leq 2\,S_{\beta,n}^\star \leq 2\,S_\beta(\Xb_n) = 4\,P_\beta(\Xb_{n+1})$.

Finally, we show that $1\leq \rho_\beta(\Xb_n) \leq 2$. Since $\SX$ is convex, for any $\Xb_n$, $n\geq 2$, the $n$ balls $\SB(\xb_i,P_\beta(\Xb_n))$ do not cover $\SX^-(P_\beta(\Xb_n)/\beta)$, which implies the existence of a $\xb\in\SX^-(P_\beta(\Xb_n)/\beta)\setminus \cup_{i=1}^n \SB(\xb_i,P_\beta(\Xb_n))$. It satisfies $d(\xb,\partial\SX) \geq P_\beta(\Xb_n)/\beta$ and $d(\xb,\Xb_n)\geq P_\beta(\Xb_n)/\beta$; therefore,
$D_\beta(\xb,\Xb_n) \geq P_\beta(\Xb_n)$ and $S_\beta(\Xb_n)\geq P_\beta(\Xb_n)$, so that $\rho_\beta(\Xb_n)\geq 1$. Any design obtained with the greedy algorithm satisfies $S_\beta(\Xb_{n+1})\leq S_\beta(\Xb_n) = 2\, P_\beta(\Xb_{n+1})$, $n\geq 1$, so that $\rho_\beta(\Xb_{n+1}) \leq 2$.
\carre

\paragraph{Minimization of a relaxed version of $\CR(\Zb_n)$.}

In \cite{PZ2019-JSC}, the following relaxed version of the coverage criterion of \cite{RoyleN98} is considered:
\be\label{relaxed-CR}
\Psi_q(\Zb_n) = \Psi_q(\Zb_n;\mu) = \left[\int_\SX \left(\frac1n\ \sum_{i=1}^n \|\zb_i - \xb\|^{-q} \right)^{-1} \,\mu(\dd\xb)\right]^{1/q}\,, \ q\neq 0\,,
\ee
with $\mu$ the uniform probability measure on $\SX$.
For any $n$-point design $\Zb_n$, it satisfies $\Psi_q(\Zb_n) \ra \CR(\Zb_n)$ as $q\ra\infty$. If $\xi_n=(1/n)\,\sum_{i=1}^n \delta_{\zb_i}$ denotes the empirical measure associated with $\Zb_n$, we can define $\psi_q(\xi_n)=\Psi_d(\Zb_n)$ and more generally
\bea 
\psi_q(\xi) = \psi_q(\xi;\mu) = \left[\int_\SX \left( \int_\SX \|\zb - \xb\|^{-q}\, \xi(\dd\zb) \right)^{-1} \,\mu(\dd\xb)\right]^{1/q}\,, \ q\neq 0\,,
\eea
for any probability measure $\xi$ on $\SX$. It is shown in \cite{PZ2019-JSC} that $\psi_q^q(\cdot)$ is a convex functional for $q>0$ (strictly convex for $q\in(0,d)$); for reasons explained in that paper, $q$ should be taken larger that $\max\{0,d-2\}$ and we use $q=d$ in the examples below. The criterion $\psi_q(\xi)$ can be minimized by the conditional (or constrained) gradient method of \cite{FrankW56}; when the step size at iteration $k$ equals $1/(k+1)$, the method corresponds to Wynn's Vertex-Direction (VD) method \cite{Wynn70} of approximate design theory (in practice, a discrete measure $\mu_Q$ supported on $\SX_Q\subset\SX$ is substituted for $\mu$, and the minimization of $\psi_q(\xi)$ corresponds to an $A$-optimal design problem). Taking the initial measure $\xi^{(1)}$ as the one-point measure supported at $\zb_1\in\SX$ and using iteratively $\xi^{(k+1)}=[k/(k+1)]\,\xi^{(k)} + \delta_{\zb_{k+1}}/(k+1)$, then, for each $n\geq 1$, $\xi^{(n)}$ has $n$ support points which define an incremental design $\Xb_n$. Here, $\zb_1 \in \Arg\min_{\zb\in\SX} \int_\SX \|\zb - \xb\|^q\, \mu(\dd\zb)$ and $\zb_{k+1}$ minimizes the directional derivative of $\psi_q^q(\cdot)$ at $\xi^{(k)}$ in the direction of the delta measure $\delta_{\zb}$ with respect to $\zb$. When a finite candidate set $\SX_C$ is used, this corresponds to
\bea
\zb_{k+1} \in\Arg\max_{\zb\in\SX_C} \int_\SX \left[ \|\zb - \xb\|^{-q} \left( \int_\SX \|\yb - \xb\|^{-q}\, \xi^{(k)}(\dd\yb) \right)^{-2} \right] \,\mu(\dd\xb) \,;
\eea
see \cite{PZ2019-JSC}. We shall denote by $\Xb_n^{VD}$ the designs constructed in this manner, with $\mu_Q$ substituted for $\mu$.

Straightforward calculations indicate that, for $q>0$, $(1/n)\, \Psi_q^q(\Zb_n,\mu_Q)$, a discrete un-normalized form of $\Psi_q^q(\Zb_n;\mu)$ defined in \eqref{relaxed-CR}, is non-increasing and supermodular (i.e., $-(1/n)\, \Psi_q^q(\Zb_n,\mu_Q)$ is submodular). Theorem~\ref{th:NemhauserWF} thus applies to the greedy maximization of $-(1/n)\, \Psi_q^q(\Zb_n,\mu_Q)$; we shall denote by $\Xb_n^{RD}$ a design obtained by greedy minimization of $\Psi_q(\Zb_n,\mu_Q)$, a Relaxed and Discretized version of $\CR(\Zb_n)$.
The lazy-greedy algorithm of Section~\ref{S:lazy-greedy} can also be applied, see Table~\ref{Tb:CompTime-A-VR-RD}.

\section*{Declarations}

\subsection*{Funding}
This work was partly supported by project INDEX (INcremental Design of EXperiments) ANR-18-CE91-0007 of the French National Research Agency (ANR).

\subsection*{Conflict of interest}
On behalf of all authors, the corresponding author states that there are no conflicts of interest.

\subsection*{Code availability}
The matlab scripts implementing the greedy and lazy greedy algorithms for the maximization of the covering criterion $\IBq$ are available at
\href{https://sdb3.i3s.unice.fr/anrindex/fr/node/5}{https://sdb3.i3s.unice.fr/anrindex}.

\bibliographystyle{spmpsci}      


\end{document}